\pdfoutput=1
\documentclass[acmtog, screen, authorversion]{acmart}

\usepackage{acronym}
\usepackage{booktabs}
\usepackage{epstopdf}
\usepackage{amsmath}
\usepackage{algorithm}
\usepackage[noend]{algpseudocode}
\usepackage[tight,footnotesize]{subfigure}
\usepackage{url}
\usepackage{tabularx}
\usepackage{color}

\acmJournal{TOMACS}
\acmVolume{0}
\acmNumber{0}
\acmArticle{00}
\acmYear{0}
\acmMonth{0}
\copyrightyear{0}
\setcopyright{acmcopyright}
\acmDOI{0000000.0000000}

\newcommand{\OR}{\ensuremath{\vee}}
\newcommand{\AND}{\ensuremath{\wedge}}

\newcommand{\circlenum}[1]{\raisebox{.5pt}{\textcircled{\raisebox{-.9pt} {\sf #1}}}}

\begin{document}

\title[Parallel Data Distribution Management]{Parallel Data Distribution Management on Shared-Memory Multiprocessors\footnotemark}

\author{Moreno Marzolla}
\orcid{0000-0002-2151-5287}
\affiliation{%
  \institution{University of Bologna}
  \department{Department of Computer Science and Engineering (DISI)}
  \streetaddress{Mura Anteo Zamboni 7}
  \city{Bologna}
  \postcode{40126}
  \country{Italy}}
\email{moreno.marzolla@unibo.it}

\author{Gabriele D'Angelo}
\orcid{0000-0002-3690-6651}
\affiliation{%
  \institution{University of Bologna}
  \department{Department of Computer Science and Engineering (DISI)}
  \streetaddress{Mura Anteo Zamboni 7}
  \city{Bologna}
  \postcode{40126}
  \country{Italy}}
\email{g.dangelo@unibo.it}

\begin{abstract}
The problem of identifying intersections between two sets
of~$d$-dimensional axis-parallel rectangles appears frequently in the
context of agent-based simulation studies. For this reason, the~High
Level Architecture~(HLA) specification -- a standard framework for
interoperability among simulators -- includes a Data Distribution
Management~(DDM) service whose responsibility is to report all
intersections between a set of subscription and update regions. The
algorithms at the core of the~DDM service are CPU-intensive, and could
greatly benefit from the large computing power of modern multi-core
processors. In this paper we propose two parallel solutions to the~DDM
problem that can operate effectively on shared-memory
multiprocessors. The first solution is based on a data structure (the
Interval Tree) that allows concurrent computation of intersections
between subscription and update regions. The second solution is based
on a novel parallel extension of the Sort Based Matching algorithm,
whose sequential version is considered among the most efficient
solutions to the~DDM problem. Extensive experimental evaluation of the
proposed algorithms confirm their effectiveness on taking advantage of
multiple execution units in a shared-memory architecture.
\end{abstract}

\begin{CCSXML}
<ccs2012>
<concept>
<concept_id>10010147.10010341.10010349.10010362</concept_id>
<concept_desc>Computing methodologies~Massively parallel and high-performance simulations</concept_desc>
<concept_significance>500</concept_significance>
</concept>
<concept>
<concept_id>10010147.10010169.10010170.10010171</concept_id>
<concept_desc>Computing methodologies~Shared memory algorithms</concept_desc>
<concept_significance>300</concept_significance>
</concept>
</ccs2012>
\end{CCSXML}

\ccsdesc[500]{Computing methodologies~Massively parallel and high-performance simulations}
\ccsdesc[300]{Computing methodologies~Shared memory algorithms}

\keywords{Data Distribution Management (DDM), Parallel And Distributed Simulation (PADS), High Level Architecture (HLA), Parallel Algorithms}

\thanks{A preliminary version of this paper appeared in the
  proceedings of the 21st International Symposium on Distributed
  Simulation and Real Time Applications (DS-RT 2017). This paper is an
  extensively revised and extended version of the previous work in
  which more than 30\% is new material.}

\maketitle

\footnotetext{
\textbf{{\color{red} This is the author's version of the article: ``Moreno Marzolla, Gabriele D'Angelo. Parallel Data Distribution Management on Shared-Memory Multiprocessors. ACM Transactions on Modeling and Computer Simulation (ACM TOMACS), Vol. 30, No. 1, Article 5. ACM, February 2020. ISSN: 1049-3301.''. The publisher version of this paper is available at \url{https://dl.acm.org/doi/10.1145/3369759}.}}}

%%%%%%%%%%%%%%%%%%%%%%%%%%%%%%%%%%%%%%%%%%%%%%%%%%%%%%%%%%%%%%%%%%%%%%%%%%%%%%
\section{Introduction}\label{sec:introduction}

Large agent-based simulations are used in many different areas such
human mobility modeling~\cite{Klugl:2007}, transportation and
logistics~\cite{Cetin:2002}, or complex biological
systems~\cite{Abraham:2015, Jacob:2004}. While there exist
recommendations and best practices for designing credible simulation
studies~\cite{Law:2009}, taming the complexity of large models remains
challenging, due to the potentially huge number of virtual entities
that need to be orchestrated and the correspondingly large amount of
computational resources required to execute the model.

The~\ac{HLA} has been introduced to partially address the problems
above. The~\ac{HLA} is a general architecture for the interoperability
of simulators~\cite{HLA} that allow users to build large models
through composition of specialized simulators, called \emph{federates}
according to the~\ac{HLA} terminology. The federates interact using a
standard interface provided by a component
called~\ac{RTI}~\cite{HLA-interf}. The structure and semantics of the
data exchanged among the federates are formalized in the~\acf{OMT}
specification~\cite{HLA-omt}.

Federates can notify events to other federates, for example to signal
a local status update that might impact other simulators. Since
notifications might produce a significant overhead in terms of network
traffic and processor usage at the receiving end, the~\ac{RTI}
provides a~\ac{DDM} service whose purpose is to allow federates to
specify which notifications they are interested in. This is achieved
through a spatial public-subscribe system, where events are associated
with an axis-parallel, rectangular region in $d$-dimensional space,
and federates can signal the~\ac{RTI} to only receive notifications
that overlap one or more subscription regions of interest.

More specifically, \ac{HLA} allows the simulation model to define a
set of \emph{dimensions}, each dimension being a range of integer
values from~$0$ to a user-defined upper bound. Dimensions may be used
as Cartesian coordinates for mapping the position of agents in 2-D or
3-D space, although the~\ac{HLA} specification does not mandate
this. A~\emph{range} is a half-open interval $[\textit{lower bound},
  \textit{upper bound})$ of values on one dimension. A \emph{region
    specification} is a set of ranges, and can be used by federates to
  denote the ``area of influence'' of status update
  notifications. Federates can signal the~\ac{RTI} the regions from
  which update notifications are to be received (\emph{subscription
    regions}). Each update notification is associated with a
  \emph{update region}: the~\ac{DDM} service identifies the set of
  overlapping subscription regions, so that the update message are
  sent only to federates owning the relevant subscription.

\begin{figure*}
  \centering\includegraphics[scale=.7]{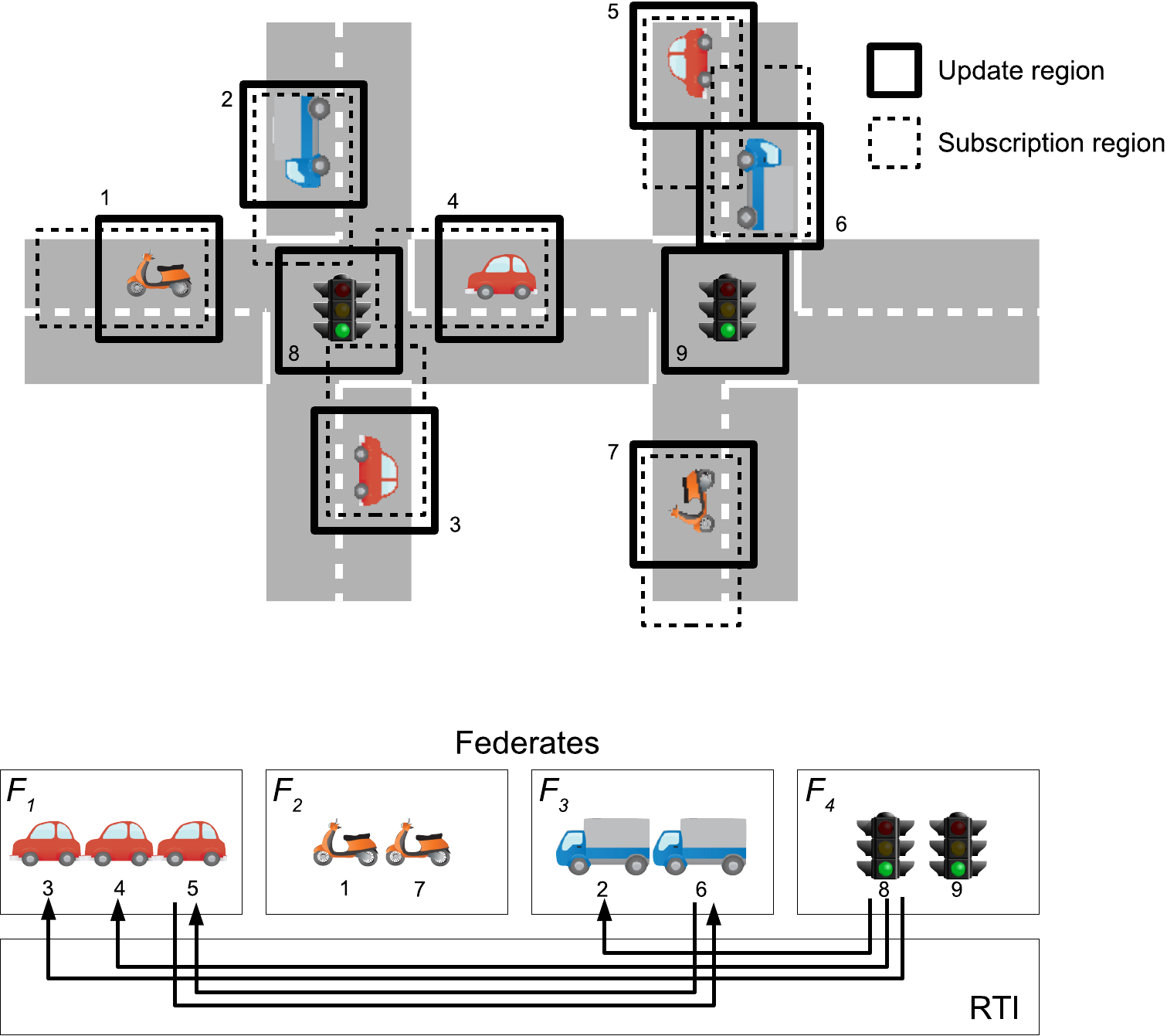}
  \caption{(Top)~Road traffic simulation example. (Bottom)~A possible
    mapping of simulated entities (vehicles and traffic lights) to
    federates.}\label{fig:road-traffic-ex}
\end{figure*}

As an example, let us consider the simple road traffic simulation
model shown in Figure~\ref{fig:road-traffic-ex} consisting of vehicles
managing intersections. Vehicles need to react both to changes of the
color of traffic lights and also to other vehicles in front of them.
This can be achieved using suitably defined update and subscription
regions. Specifically, each vehicle is enclosed within an update
region (thick box) and a subscription region (dashed box), while each
traffic light is enclosed within an update region only. A traffic
light is a pure generator of update notifications, while vehicles are
both producers and consumers of events. Each vehicle generates
notifications to signal a change in its position, and consumes events
generated by nearby vehicles and traffic lights. We assume that a
vehicle can safely ignore what happens behind it; therefore,
subscription regions are skewed towards the direction of motion.

If the scenario above is realized through an~\ac{HLA}-compliant
simulator, entities need to be assigned to federates. We suppose that
there are four federates, $F_1$ to~$F_4$. $F_1$, $F_2$ and $F_3$
handle cars, scooters and trucks respectively, while $F_4$ manages the
traffic lights. Each simulated entity registers subscription and
update regions with the~\ac{RTI}; the~\ac{DDM} service can then match
subscription and update regions, so that update notifications can be
sent to interested entities. In our scenario, vehicles~$2$, $3$
and~$4$ receive notifications from the traffic light~$8$; vehicles~$5$
and~$6$ send notifications to each other, since their subscription and
update regions overlap. The communication pattern between federates is
shown in the bottom part of Figure~\ref{fig:road-traffic-ex}.

As can be seen, at the core of the~\ac{DDM} service there is an
algorithm that solves an instance of the general problem of reporting
all pairs of intersecting axis-parallel rectangles in a
$d$-dimensional space. In the context of~\ac{DDM}, \emph{reporting}
means that each overlapping subscription-update pair must be reported
exactly once, without any implied ordering. This problem is well known
in computational geometry, and can be solved either using appropriate
algorithms and/or ad-hoc spatial data structures as will be discussed
in Section~\ref{sec:region-matching}. However, it turns out that
\ac{DDM} implementations tend to rely on less efficient but simpler
solutions. The reason is that spatial data structures can be quite
complicated, and therefore their manipulation may require a
significant overhead that might be not evident from their asymptotic
complexity.

The increasingly large size of agent-based simulations is posing a
challenge to the existing implementations of the \ac{DDM} service. As
the number of regions increases, so does the execution time of the
intersection-finding algorithms. A possible solution comes from the
computer architectures domain. The current trend in microprocessor
design is to put more execution units (cores) in the same processor;
the result is that multi-core processors are now ubiquitous, so it
makes sense to try to exploit the increased computational power to
speed up the~\ac{DDM} service~\cite{gda-simpat}. Therefore, an obvious
parallelization strategy for the intersection-finding problem is to
distribute the rectangles across the processor cores, so that each
core can work on a smaller problem.

\begin{figure*}
  \centering\includegraphics[scale=.7]{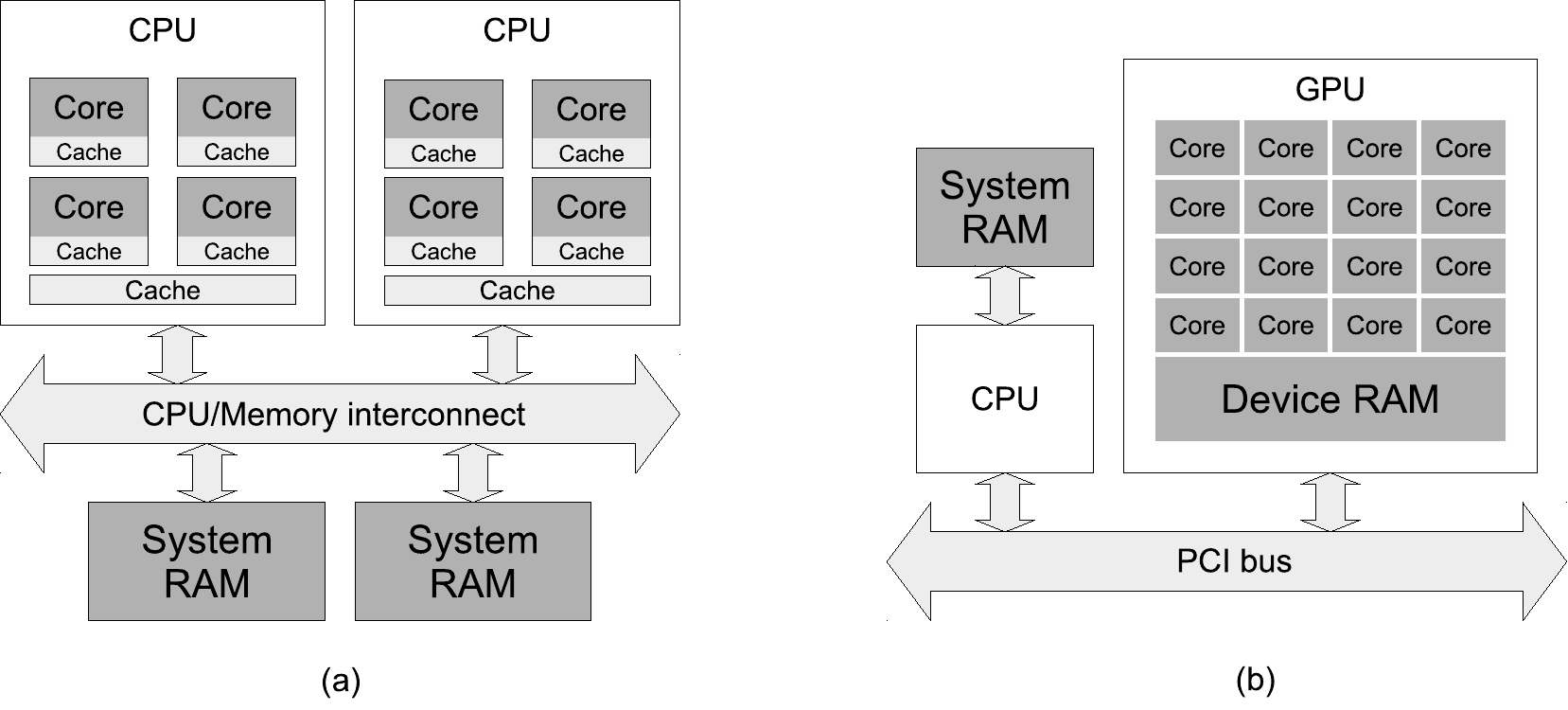}
  \caption{A shared-memory multiprocessor
    system.}\label{fig:shared-memory}
\end{figure*}

Shared-memory multiprocessors are a family of parallel systems where
multiple processors share one or more blocks of Random Access
Memory~(RAM) through an interconnection network
(Figure~\ref{fig:shared-memory}~(a)). A modern multi-core CPU contains
multiple independent execution units (cores), that are essentially
stand-alone processors. Cache hierarchies within each core, and shared
by all cores of the same~CPU, are used to mitigate the memory access
bottleneck, known as the \emph{memory wall}~\cite{Wulf:1995}.

General-Purpose GPU computing is another realization of the multi-core
paradigm. \acp{GPU} were originally intended as specialized devices
for producing graphical output, but have now evolved into
general-purpose parallel co-processors. A high-level overview of
a~\ac{GPU} is shown in Figure~\ref{fig:shared-memory}~(b). A~\ac{GPU}
includes a large number of cores that share a common memory, called
\emph{device memory}. The device memory is physically separate from
the main system~RAM, so that programs and data must be transferred
from system~RAM to device memory before the~\ac{GPU} can start
execution. At the end of the computation, results need to be
transferred back to system~RAM. While a single~\ac{GPU} core is less
powerful than a~CPU core, a~\ac{GPU} has more cores than a
typical~CPU, and therefore provides a higher aggregate
throughput. However, this comes at a cost: \ac{GPU} programming is in
general more complex than programming a multicore~CPU; additionally,
CPUs support more memory than~\acp{GPU}, meaning that CPU-GPU
communication might be a bottleneck when processing large datasets.
Moreover, \acp{GPU} are based on the~\ac{SIMD} paradigm, where a
single instruction stream is executed on multiple data items. This
paradigm is well suited for applications with regular data access
pattern (e.g., linear algebra). Applications with conditional branches
or irregular data access patterns may require considerable more effort
to be implemented efficiently, but are nevertheless possible: for
example, non-trivial but efficient~\ac{GPU} implementations of the
Time Warp optimistic synchronization protocol~\cite{Liu:2017} and of
the Bellman-Ford shortest path algorithm~\cite{Busato:2016} have been
realized, despite the fact that both fall outside the these
application exhibits regular data access patterns. Finally, it must be
observed that general-purpose~\acp{GPU} are currently not as
ubiquitous as multicore~CPUs, since they are add-on cards that must be
purchased separately.

Shared-memory multiprocessors are interesting for several reasons.
first, they are ubiquitous since they power virtually everything from
smartphones and single-board computers up to high performance
computing systems. Moreover, shared-memory architectures are in
general simpler to program than distributed-memory architectures,
since the latter require explicit message exchanges to share data
between processors. Support for shared-memory programming has been
added to traditional programming languages such as C, C++ and
FORTRAN~\cite{OpenMP}, further reducing the effort needed to write
parallel applications.

Unfortunately, writing \emph{efficient} parallel programs is not easy.
Many serial algorithms can not be made parallel by means of simple
transformations. Instead, new parallel algorithms must be designed
from scratch around the features of the underlying execution
platform. The lack of abstraction of parallel programming is due to
the fact that the user must leverage the strengths (and avoid the
weaknesses) of the underlying execution platform in order to get the
maximum performance. The result is that, while parallel architectures
are ubiquitous -- especially shared-memory ones -- parallel programs
are not, depriving users from a potential performance boost on some
classes of applications.

In this paper we present two solutions to the~\ac{DDM} problem that
are suitable for shared-memory multiprocessors. The first solution,
called~\ac{ITM}, is based on the \emph{interval tree} data structure,
that represents subscription or update regions in such a way that
intersections can be computed in parallel. The second solution,
called~\emph{Parallel \acs{SBM}}, is a parallel version
of~\ac{SBM}~\cite{Raczy2005}, a state-of-the-art implementation of
the~\ac{DDM} service.

This paper is organized as follows. In
Section~\ref{sec:region-matching}, we review the scientific literature
related to the~\ac{DDM} service and describe in detail some of the
existing~\ac{DDM} algorithms that will be later used in the paper. In
Section~\ref{sec:interval-tree} we describe the interval tree data
structure and the~\ac{ITM} parallel matching algorithm. In
Section~\ref{sec:parallel-sort-matching}, we present the main
contribution of this work, i.e., a parallel version of the~\ac{SBM}
algorithm. In Section~\ref{sec:experimental-evaluation} we
experimentally evaluate the performance of parallel~\ac{SBM} on two
multi-core processors. Finally, the conclusions will be discussed in
Section~\ref{sec:conclusions}.

%%%%%%%%%%%%%%%%%%%%%%%%%%%%%%%%%%%%%%%%%%%%%%%%%%%%%%%%%%%%%%%%%%%%%%%%%%%%%%
\section{Problem statement and existing solutions}\label{sec:region-matching}

In this paper we address the region matching problem defined as
follows:\medskip

\noindent\textbf{Region Matching Problem.} Given two sets $\mathbf{S}
= \{S_1, \ldots, S_n\}$ and $\mathbf{U}=\{U_1, \ldots, U_m\}$ of
$d$-dimensional, axis-parallel rectangles (also called $d$-rectangles),
enumerate all pairs $(S_i, U_j) \subseteq \mathbf{S} \times
\mathbf{U}$ such that $S_i \cap U_j \neq \emptyset$; each pair must be
reported exactly once, in no particular order.\medskip

\begin{figure*}[t]
\centering\includegraphics[scale=.7]{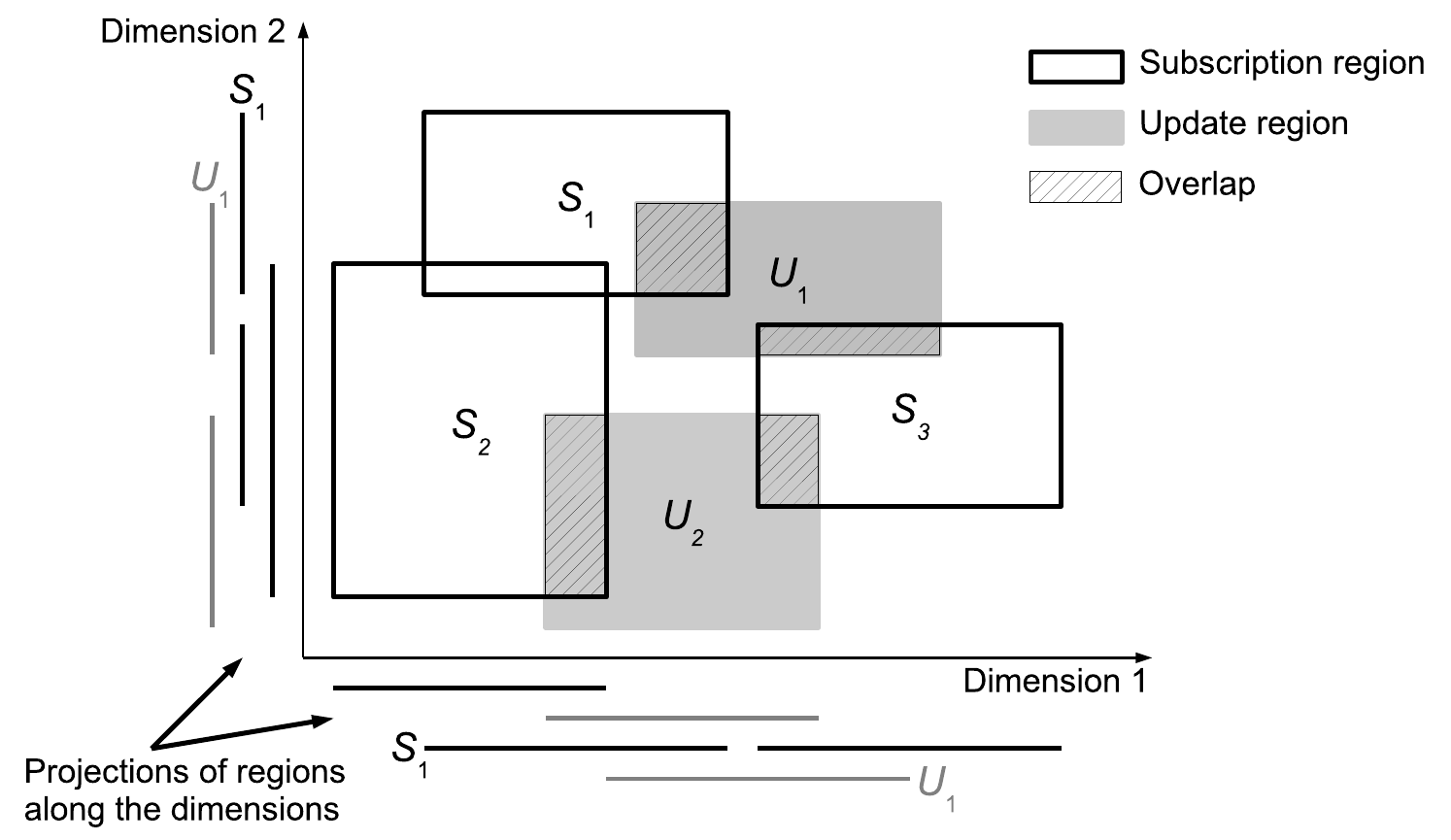}
\caption{An example of the region Matching Problem in $d=2$
  dimensions. The list of overlapping (subscription, update) pairs is
  $\{ (S_1, U_1), (S_2, U_2), (S_3, U_1), (S_3,
  U_2)\}$}\label{fig:ddm_example}
\end{figure*}

Although the~\ac{HLA} specification only allows integer coordinates,
we address the more general case in which the coordinates of the edges
of the $d$-rectangles are arbitrary real numbers.

Figure~\ref{fig:ddm_example} shows an instance of the region matching
problem in~$d=2$ dimensions with three subscription regions
$\mathbf{S} = \{S_1, S_2, S_3\}$ and two update regions $\mathbf{U} =
\{U_1, U_2\}$. In this example there are four overlapping
subscription-update pairs $\{ (S_1, U_1),\ (S_2, U_2),\ (S_3,
U_1),\ (S_3, U_2) \}$.

The time complexity of the region matching problem is
\emph{output-sensitive}, since it depends on the size of the output in
addition to the size of the input. Therefore, if there are~$K$
overlapping regions, any region matching algorithm requires
time~$\Omega(K)$. Since there can be at most $n \times m$ overlaps,
the worst-case complexity of the region matching problem is $\Omega(n
\times m)$. In practice, however, the number of intersections is much
smaller than $n \times m$.

\begin{algorithm}[t]
\caption{\textsc{Intersect-1D}$(x, y)$}\label{alg:intersect1d}
\begin{algorithmic}
\State \textbf{return} $x.\emph{low} \leq y.\emph{high} \AND\ y.\emph{low} \leq x.\emph{high}$
\end{algorithmic}
\end{algorithm}

One of the key steps of any matching algorithm is testing whether two
$d$-rectangles overlap. The case $d=1$ is very simple, as it reduces
to testing whether two half-open intervals
$x=[x.\textit{low},x.\textit{high})$, $y=[y.\textit{low},
    y.\textit{high})$ intersect; this happens if and only if
\[
x.\emph{low} < y.\emph{high} \AND\ y.\emph{low} < x.\emph{high}
\]
\noindent (see Algorithm~\ref{alg:intersect1d}).

The case~$d>1$ can be reduced to the case~$d=1$ by observing that two
$d$-rectangles overlap if and only if their projections along each
dimension overlap. For example, looking again at
Figure~\ref{fig:ddm_example} we see that the projections of~$U_1$
and~$S_1$ overlap on both dimensions, so we can conclude that the
regions~$U_1$ and~$S_1$ intersect. Therefore, any algorithm that
solves the region matching problem for two sets of~$n$ and~$m$
segments in time~$O\left(f(n,m)\right)$ can be extended to an~$O\left(
d \times f(n,m) \right)$ algorithm for the~$d$-dimensional case, by
executing the 1-D algorithm on each dimension and computing the
intersection of the partial results\footnote{The $O(\left(d \times
  f(n,m) \right)$ bound holds provided that combining the partial
  results can be done in time $O\left( f(n,m) \right)$; this is indeed
  the case for any reasonable $f(n,m)$ using hash-based set
  implementations, as we will discuss in
  Section~\ref{sec:experimental-evaluation}.}. Since the parameter~$d$
is fixed for every problem instance, and much smaller than~$n$ or~$m$,
it can be treated as a constant so we get $O\left( d \times f(n,m)
\right) \subset O\left( f(n,m) \right)$. Therefore, solving the
general case is, under reasonable circumstances, asymptotically not
harder than solving the special case $d=1$.

In the rest of this section we provide some background on the region
matching problem. We report some known lower bounds for specific
formulations of this problem; we then review some algorithms and data
structures that have been developed in the context of computational
geometry research. Finally, we describe in details some relevant
solutions developed within the~\ac{HLA} research community: \ac{BFM},
\ac{GBM}, and \acf{SBM}.  Since we will frequently refer to these
algorithms in this paper, we provide all the necessary details below.
A comprehensive review of the $d$-dimensional region matching problem
is outside the scope of this work, and has already been carried out by
Liu and Theodoropoulos~\cite{Liu:2014} to which the interested reader
is referred.

\paragraph{Lower Bounds}
Over time, several slightly different formulations of the region
matching problem have been considered. The most common formulation is
to find all overlapping pairs among a set of~$N$ rectangles, without
any distinction between subscription and update regions. One of the
first efficient solutions for this problem is due to Bentley and
Wood~\cite{Bentley:1980} who proposed an algorithm for the
two-dimensional case requiring time $\Theta(N \lg N + K)$, where~$K$
is the cardinality of the result. They proved that the result is
optimal by showing a lower bound $\Omega(N \lg N + K)$ for this type
of problem.

Petty and Morse~\cite{Petty:2004} studied the computational complexity
of the following formulation of the region matching problem: given an
initially empty set~$R$ of $d$-dimensional regions, apply a sequence
of~$N$ operations of any of the following types: (\emph{i})~insert a
new region in~$R$; (\emph{ii})~delete an existing region from~$R$;
(\emph{iii})~enumerate all regions in~$R$ overlapping with a given
test region. They showed that a lower bound on the computational
complexity of this problem is $\Omega(N \lg N)$ by reduction to
binary search, and an upper bound is $O(N^2)$; the upper bound can be
achieved with the Brute Force algorithm that we will describe shortly.

\paragraph{Computational Geometry approaches}
The problem of counting or enumerating intersections among
$d$-dimensional rectangles is of great importance in computational
geometry, and as such received considerable attention. Existing
approaches rely on a combination of algorithmic techniques such as
sorting, searching and partitioning~\cite{Six:1980, Six:1982,
  Bentley:1980, Edelsbrunner:1983}, and data structures used to speed
up various types of spatial queries (containment, overlap, counting,
filtering).

The aforementioned algorithm by Bentley and Wood~\cite{Bentley:1980}
can report all intersections within a set of~$N$ rectangles in time
$\Theta(N \lg N + K)$. The algorithm is quite complex and operates in
two phases: the first phase takes care of rectangles whose sides
intersect using the algorithm described in~\cite{Bentley:1979}; the
second phase takes care of those rectangles which are contained into
another one, using a data structure called \emph{segment tree}. While
the algorithm is optimal, it cannot be generalized for the case $d>2$
and its implementation is nontrivial.

A simpler algorithm for the case~$d=2$ has been discovered by Six and
Wood~\cite{Six:1980}. The algorithm works by sorting the endpoints of
the rectangles along one of the dimensions, and then scanning the
endpoints updating an auxiliary data structure called \emph{interval
  tree} (which, despite the similar name, is not related to the
interval tree that will be described in
Section~\ref{sec:interval-tree}). The algorithm by Six and Wood runs
in time $\Theta(N \lg N + K)$, but is much easier to implement than
Bentley and Wood's. A generalization for the case $d>2$ has been
described in~\cite{Six:1982} and requires time $O\left(2^{d-1} n
\lg^{d-1} n + K\right)$ and space $O\left(2^{d-1} n \lg^{d-1}
n\right)$.

Edelsbrunner~\cite{Edelsbrunner:1983} proposed an improved algorithm
based on a \emph{rectangle tree} data structure that can report all
intersections among a set of~$N$ $d$-rectangles in time $O\left(N
\lg^{2d-3} N + K\right)$ and space $O\left(N \lg^{d-2} N\right)$,
thus improving the previous results.

Spatial data structures for the rectangle intersection problem include
the $k$-$d$ tree~\cite{Rosenberg1985}, the
quad-tree~\cite{Finkel:1974}, the R-tree~\cite{Guttman1984} and the~BSP
tree~\cite{Naylor:1990}. These data structures use various types of
hierarchical spatial decomposition techniques to store volumetric
objects. Spatial data structures are widely used in the context of
geographical information systems, since they allow efficient execution
of range queries, e.g., reporting all objects inside a user-specified
range. However, some of these data structures have been adapted to
solve the~\ac{DDM} problem. For example, Eroglu et
al.~\cite{Eroglu:2008} use a quad-tree to improve the grid-based
matching algorithm used in~\ac{HLA} implementations.

In~\cite{Ahn:2012} the authors propose a binary partition-based
matching algorithm whose aim is to reduce the number of overlap tests
that need to be performed between subscription and update regions.
Experimental evaluation shows that the algorithm works well in some
settings, but suffers from a worst case cost of $O(N^2 \lg N)$
where~$N$ is the total number of subscription and update regions.

Geometric algorithms and data structures have the drawback of being
quite complex to implement and, in many real-world situations, slower
than less efficient but simpler solutions. For example, Petty and
Mukherjee~\cite{Petty:1997} showed that a simple grid-based matching
algorithm performs faster than the $d$-dimensional quad-tree variant
from~\cite{VanHook:1996}.  Similarly, Devai and
Neumann~\cite{Devai2010} propose a rectangle-intersection algorithm
that is implemented using only arrays and that can enumerate all~$K$
intersections among~$N$ rectangles in time~$O(N \lg N + K)$ time
and~$O(N)$ space.

\paragraph{Brute-Force Matching}

\begin{algorithm}[t]
\caption{\textsc{BruteForce-1D}$(\mathbf{S}, \mathbf{U})$}\label{alg:bfm}
\begin{algorithmic}[1]
\ForAll{subscription intervals $s \in \mathbf{S}$}\label{alg:bfm:for}
\ForAll{update intervals $u \in \mathbf{U}$}
\If {\Call{Intersect-1D}{$s, u$}}\label{alg:bfm:body}
\Call{Report}{$s, u$}
\EndIf
\EndFor
\EndFor\label{alg:bfm:endfor}
\end{algorithmic}
\end{algorithm}

The simplest solution to the segment intersection problem is
the~\ac{BFM} approach, also called Region-Based matching
(Algorithm~\ref{alg:bfm}). The~\ac{BFM} algorithm, as the name
suggests, checks all $n \times m$ subscription-update pairs $(s, u)$
and reports every intersection by calling a model-specific function
\Call{Report}{$s$, $u$} whose details are not shown.

The~\ac{BFM} algorithm requires time~$\Theta(n m)$; therefore, it is
optimal only in the worst case, but very inefficient in
general. However, \ac{BFM} exhibits an embarrassingly parallel
structure since the loop iterations
(lines~\ref{alg:bfm:for}--\ref{alg:bfm:endfor}) are independent.
Therefore, on a shared-memory architecture with~$P$ processors it is
possible to distribute the iterations across the processors; each
processor will then operate on a subset of $nm /P$ intervals without
the need to interact with other processors. The parallel version
of~\ac{BFM} requires time~$\Theta(nm/P)$.

\paragraph{Grid-Based Matching}

\begin{figure}[t]
\centering\includegraphics[scale=.7]{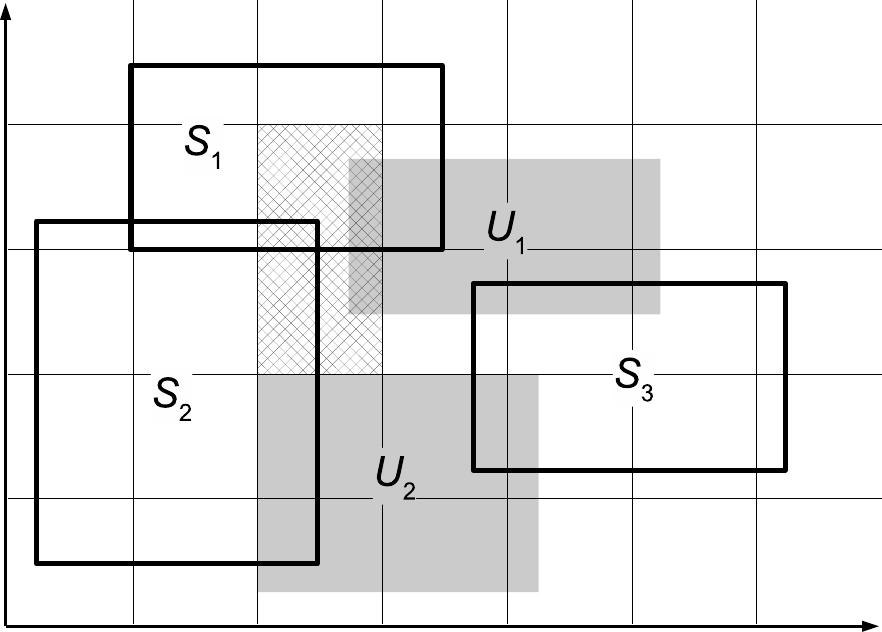}
\caption{Grid-based matching in $d=2$ dimensions.}\label{fig:ddm_example_grid}
\end{figure}

\begin{algorithm}[t]
\caption{\textsc{Grid-1D}$(\mathbf{S}, \mathbf{U})$}\label{alg:gbm}
\begin{algorithmic}[1]
  \Require{$\mathit{ncells}\ \textrm{number of grid cells}$}
  \State $G \gets \textrm{array}[0..\mathit{ncells}-1]\ \textrm{of empty lists}$
  \State $\mathit{lb} \gets \textrm{minimum of the lower bounds of all intervals in $\mathbf{S} \cup \mathbf{U}$}$
  \State $\mathit{ub} \gets \textrm{maximum of the upper bounds of all intervals in $\mathbf{S} \cup \mathbf{U}$}$
  \State $\mathit{width} \gets (\mathit{ub} - \mathit{lb}) / \mathit{ncells}$
  \ForAll{update regions $u \in \mathbf{U}$}\Comment{Build the grid}\label{alg:gbm:first-begin}
  \State $i \gets \lfloor (u.\mathit{lower} - \mathit{lb}) / \mathit{width} \rfloor$
  \While{$(i < \mathit{ncells})\ \AND\ (i \times \mathit{width} < u.\mathit{upper})$}
  \State Add $u$ to the list $G[i]$\label{alg:gbm:list-add}
  \State $i \gets i + 1$
  \EndWhile
  \EndFor\label{alg:gbm:first-end}
  \State $\mathit{res} \gets \emptyset$
  \ForAll{subscription regions $s \in \mathbf{S}$}\Comment{Find intersections}\label{alg:gbm:second-begin}
  \State $i \gets \lfloor (s.\mathit{lower} - \mathit{lb}) / \mathit{width} \rfloor$
  \While{$(i < \mathit{ncells})\ \AND\ (i \times \mathit{width} < s.\mathit{upper})$}
  \ForAll{update regions $u \in G[i]$}
  \If{$\Call{Intersect-1D}{s, u}\ \AND\ (s, u) \notin \mathit{res}$}\label{alg:gbm:check-overlap}
  \State $\mathit{res} \gets \mathit{res} \cup (s, u)$
  \State \Call{Report}{$s, u$}
  \EndIf
  \EndFor
  \State $i \gets i + 1$
  \EndWhile
  \EndFor\label{alg:gbm:second-end}
\end{algorithmic}
\end{algorithm}

The~\ac{GBM} algorithm~\cite{VanHook:1996,bou2005} improves over~\ac{BFM} by
trying to reduce the number of pairs that are checked for
overlap. \ac{GBM} works by partitioning the domain into a regular mesh
of $d$-dimensional cells. Each subscription or update region is mapped
to the grid cells it overlaps with. Events generated by an update
region~$U_j$ are sent to all subscription regions that share at least
one cell with~$U_j$. However, this could generate spurious
notifications: for example, the subscription region~$S_2$ in
Figure~\ref{fig:ddm_example_grid} shares the hatched grid cells
with~$U_1$, but does not overlap with~$U_1$. Spurious notifications
can be eliminated by testing for overlap all subscription and update
regions sharing the same grid cell. Essentially, this is equivalent to
applying the brute-force (or any other region matching)
algorithm to the regions sharing the same
grid cell~\cite{Tan:2000-2}.

Algorithm~\ref{alg:gbm} shows an implementation of~\ac{GBM} for the
case~$d=1$. The algorithm consists of two phases. During the first
phase (lines~\ref{alg:gbm:first-begin}--\ref{alg:gbm:first-end}) the
algorithm builds an array~$G$ of lists, where $G[i]$ contains the
update regions that overlap with the $i$-th grid cell. The grid cells
are determined by first computing the bounding interval $[\mathit{lb},
  \mathit{ub})$ of all regions in~$\mathbf{S}$ and~$\mathbf{U}$. Then,
  the bounding interval is evenly split into $\mathit{ncells}$
  segments of width $(\mathit{ub} - \mathit{lb})/\mathit{ncells}$ so
  that the $i$-th grid cell corresponds to the interval $[\mathit{lb}
    + i\times\mathit{width}, \mathit{lb} +
    (i+1)\times\mathit{width})$.  The parameter $\mathit{ncells}$ must
    be provided by the user.

During the second phase
(lines~\ref{alg:gbm:second-begin}--\ref{alg:gbm:second-end}), the
algorithm scans the list of subscription regions. Each subscription
region is compared with the update regions on the lists of the cells
it overlaps with (line~\ref{alg:gbm:check-overlap}). Since
subscription and update regions may span more than one grid cell, the
algorithm keeps a set $\mathit{res}$ of all intersections found so far
in order to avoid reporting the same intersection more than once.

If the regions are evenly distributed, each grid cell will contain
$n/\mathit{ncells}$ subscription and $m/\mathit{ncells}$ update
intervals on average. Therefore, function \Call{Intersect-1D}{} will
be called $O(\mathit{ncells} \times n \times m / \mathit{ncells}^2) =
O(n \times m / \mathit{ncells})$ times on average. Initialization of
the array~$G$ requires time $O(\mathit{ncells})$. The upper and lower
bounds $\mathit{lb}$ and $\mathit{ub}$ can be computed in time $O(n +
m)$. If the set $\mathit{res}$ is implemented using bit vectors,
insertions and membership tests can be done in constant time. We can
therefore conclude that the average running time of
Algorithm~\ref{alg:gbm} is $O(\mathit{ncells} + n \times m /
\mathit{ncells})$.

The average-case analysis above only holds if subscription and update
regions are uniformly distributed over the grid, which might or might
not be the case depending on the simulation model. For example, in the
presence of a localized cluster of interacting agents, it might happen
that the grid cells around the cluster have a significantly larger
number of intervals than other cells.  Additionally, the number of
cells~$\mathit{ncells}$ is a critical parameter. Tan et
al.~\cite{Tan:2000} showed that the optimal cell size depends on the
simulation model and on the execution environment, and is therefore
difficult to predict \emph{a priori}.

Observe that the iterations of the loop on
lines~\ref{alg:gbm:second-begin}--\ref{alg:gbm:second-end} are
independent and can therefore be executed in parallel. If we do the
same for the loop on
lines~\ref{alg:gbm:first-begin}--\ref{alg:gbm:first-end}, however, a
data race arises since multiple processors might concurrently update
the list~$G[i]$. This problem can be addressed by ensuring that
line~\ref{alg:gbm:list-add} is executed atomically, e.g., by enclosing
it inside a critical section or employing a suitable data structure
for the lists~$G[i]$ that supports concurrent appends. We will discuss
this in more details in Section~\ref{sec:experimental-evaluation}. Finally, it
is worth noticing that some variants of~\ac{GBM} have been proposed to 
combine the grid-based method with a region-based strategy~\cite{bou2002}.

\begin{algorithm}[t]
\caption{\textsc{Sort-Based-Matching-1D}$(\mathbf{S}, \mathbf{U})$}\label{alg:sbm}
\begin{algorithmic}[1]
\State $T \gets \emptyset$
\ForAll{regions $x \in \mathbf{S} \cup \mathbf{U}$}
\State Insert $x.\textit{lower}$ and $x.\textit{upper}$ in $T$
\EndFor
\State Sort $T$ in non-decreasing order
\State $\texttt{SubSet} \gets \emptyset$, $\texttt{UpdSet} \gets \emptyset$
\ForAll{endpoints $t \in T$ in non-decreasing order}\label{alg:sbm:loop-start}
\If{$t$ belongs to subscription region $s$}
\If{$t$ is the lower bound of $s$}
\State $\texttt{SubSet} \gets \texttt{SubSet} \cup \{s\}$
\Else
\State $\texttt{SubSet} \gets \texttt{SubSet} \setminus \{s\}$
\ForAll{$u \in \texttt{UpdSet}$} \Call{Report}{$s, u$}
\EndFor
\EndIf
\Else\Comment{$t$ belongs to update region $u$}
\If{$t$ is the lower bound of $u$}
\State $\texttt{UpdSet} \gets \texttt{UpdSet} \cup \{u\}$
\Else
\State $\texttt{UpdSet} \gets \texttt{UpdSet} \setminus \{u\}$
\ForAll{$s \in \texttt{SubSet}$} \Call{Report}{$s, u$}
\EndFor
\EndIf
\EndIf
\EndFor\label{alg:sbm:loop-end}
\end{algorithmic}
\end{algorithm}

\begin{figure*}[t]
\centerline{\includegraphics[scale=.7]{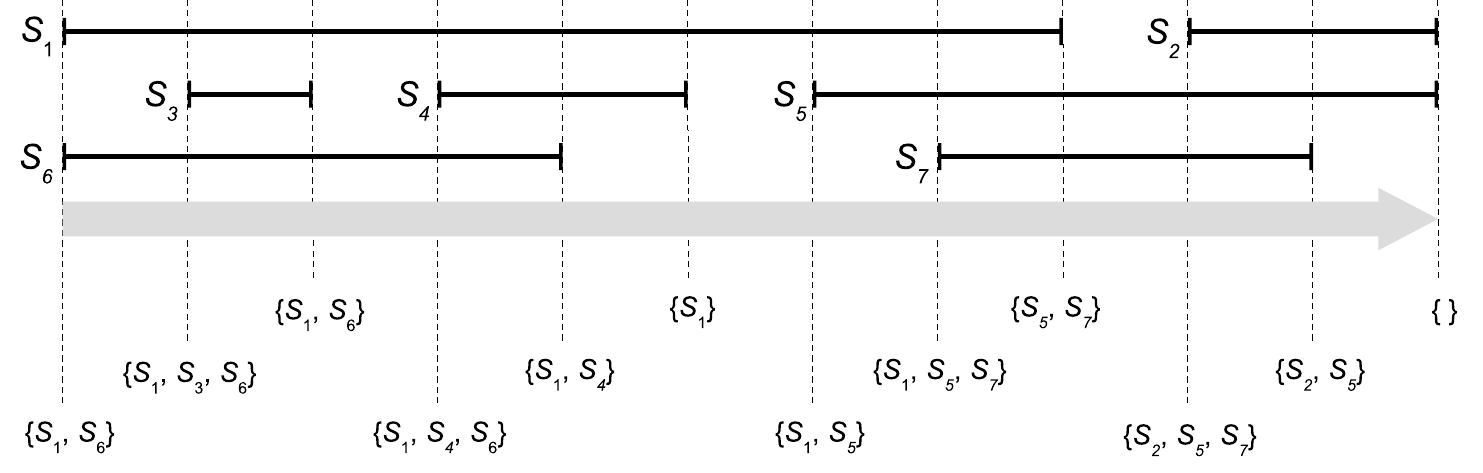}}
\caption{Value assigned by the~\ac{SBM} algorithm to the
  \texttt{SubSet} variable as the endpoints are swept from left to
  right.}\label{fig:sbm-example}
\end{figure*}

\paragraph{Sort-Based Matching}
\acl{SBM}~\cite{Jun2002,Raczy2005} is an efficient solution to the
region matching problem in $d=1$ dimensions. \ac{SBM} scans the sorted
list of endpoints of the subscription and update intervals, keeping
track of which regions are active at any point; a region is active if
the left endpoint has been scanned, but the right endpoint has
not. When the right endpoint of a subscription (resp., update)
region~$x$ is encountered, all currently active update (resp.,
subscription) regions are known to overlap with~$x$.

The~\ac{SBM} algorithm is illustrated in
Algorithm~\ref{alg:sbm}. Given~$n$ subscription intervals~$\mathbf{S}$
and~$m$ update intervals~$\mathbf{U}$, \ac{SBM} considers each of the
$2 \times (n+m)$ endpoints in non-decreasing order; two sets
\texttt{SubSet} and \texttt{UpdSet} are used to keep track of the
active subscription and update regions, respectively, at every
point~$t$.  When the upper bound of an interval~$x$ is encountered, it
is removed from the corresponding set of active regions, and the
intersections between~$x$ and every active region of the opposite kind
are reported. Observe that Algorithm~\ref{alg:sbm} never calls the
function \Call{Intersect-1D}{} to check whether two regions overlap.
Figure~\ref{fig:sbm-example} illustrates how the \texttt{SubSet}
variable is updated while~\ac{SBM} sweeps through a set of
subscription intervals (update intervals are handled in the same way).

Let~$N=n+m$ be the total number of intervals; then, the number of
endpoints is $2N$. The~\ac{SBM} algorithm uses simple data structures
and requires $O\left( N \lg N \right)$ time to sort the vector of
endpoints, plus~$O(N)$ time to scan the sorted vector. During the scan
phase, $O(K)$ time is spent to report all~$K$ intersections. The
overall computational cost of~\ac{SBM} is therefore $O\left( N \lg N +
K \right)$.

Li et al.~\cite{Li:2018} improved the~\ac{SBM} algorithm
by reducing the size of the vectors to be sorted and employing the
binary search algorithm on the (smaller) sorted vectors of
endpoints. The execution time is still dominated by the $O\left(N \lg
N\right)$ time required to sort the smaller vectors of endpoints, but
the improved algorithm is faster in practice than~\ac{SBM} due to
lower constants hidden in the asymptotic notation.

Pan et al.~\cite{Pan2011} extended~\ac{SBM} to deal with a dynamic
setting where regions are allowed to change their position or size
without the need to execute the~\ac{SBM} algorithm from scratch after
each update.

So far, no parallel version of~\ac{SBM} exists. \ac{SBM} cannot be
easily parallelized due to the presence of a sequential scan phase
that is intrinsically serial. This problem will be addressed in
Section~\ref{sec:parallel-sort-matching}, where a parallel version
of~\ac{SBM} will be described.

\paragraph*{Parallel algorithms for the region matching problem}
So far, only a few parallel algorithms for the region matching problem
have been proposed. Liu and Theodoropoulos~\cite{Liu:2009, Liu:2013}
propose a parallel region matching algorithm that partitions the
routing space into blocks, and assigns partitions to processors using
a master-worker paradigm. Each processor then computes the
intersections among the regions that overlap the received
partitions. In essence, this solution resembles a parallel version of
the~\ac{GBM} algorithm.

In~\cite{rao2013} the performance of parallel versions of~\ac{BFM} and
grid-based matching (fixed, dynamic and hierarchical) are compared.
In this case, the preliminary results presented show that the
parallel~\ac{BFM} has a limited scalability and that, in this specific
case, the hierarchical grid-based matching has the best performance.

In~\cite{Yanbing2015}, a parallel ordered-relation-based matching
algorithm is proposed. The algorithm is composed of five phases:
projection, sorting, task decomposition, internal matching and
external matching. In the experimental evaluation, a~MATLAB
implementation is compared with the sequential~\ac{SBM}. The results
show that, with a high number of regions the proposed algorithm is
faster than~\ac{SBM}.

%%%%%%%%%%%%%%%%%%%%%%%%%%%%%%%%%%%%%%%%%%%%%%%%%%%%%%%%%%%%%%%%%%%%%%%%%%%%%%
\section{Parallel Interval Tree Matching}\label{sec:interval-tree}

In this section we describe the parallel~\ac{ITM} algorithm for
solving the region matching problem in one
dimension. \ac{ITM}~\cite{gda-dsrt-2013} is based on the
\emph{interval tree} data structure. An interval tree is a balanced
search tree that stores a dynamic set of intervals; it supports
insertions, deletions, and queries to get the list of segments that
intersect a given interval~$q$.

\begin{figure*}[t]
\centering\includegraphics[scale=0.8]{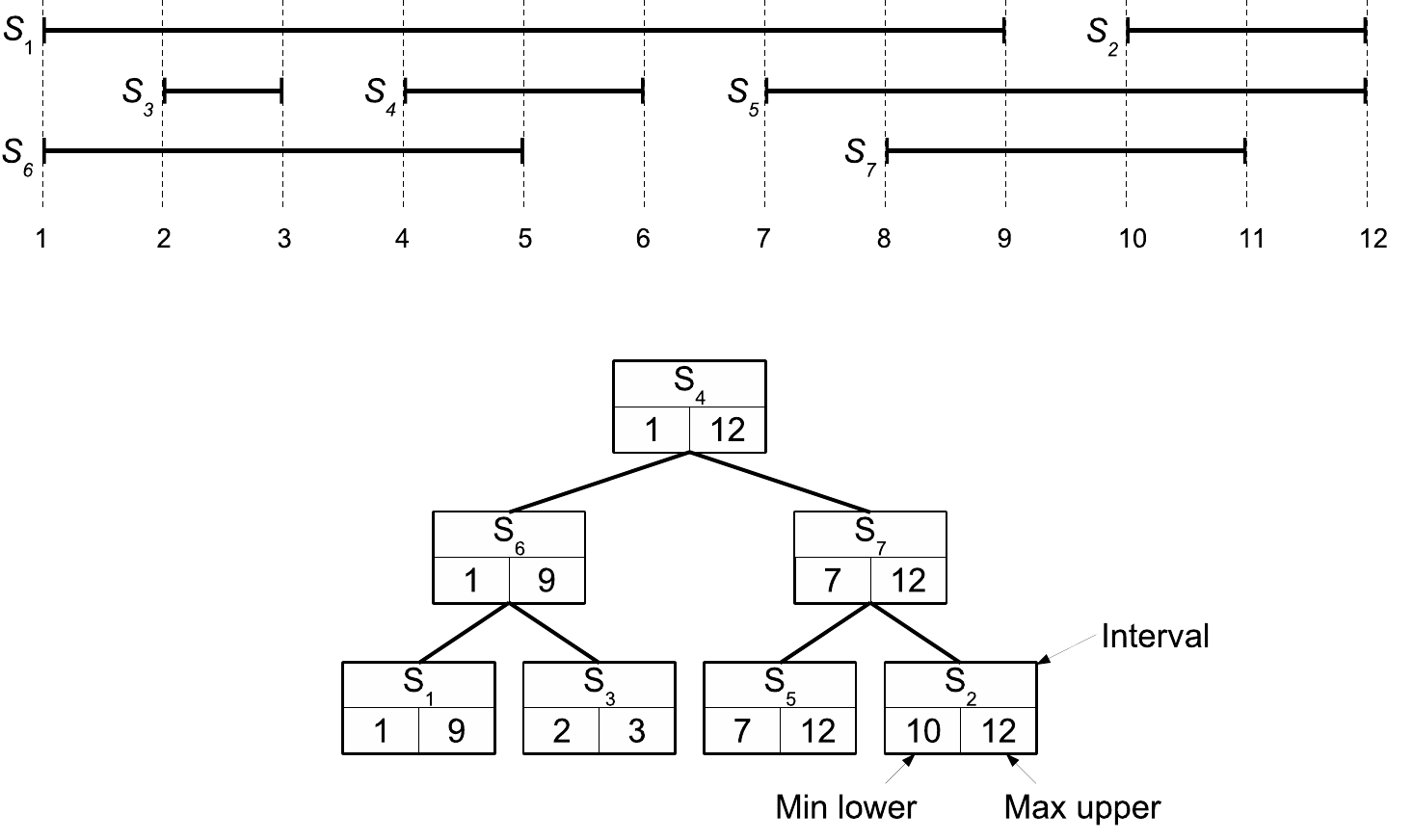}
\caption{Interval Tree representation of a set of intervals}\label{fig:interval_tree}
\end{figure*}

Figure~\ref{fig:interval_tree} shows a set of intervals and their tree
representation. The tree is realized using an augmented~AVL
tree~\cite{avl} as described in~\cite[Chapter 14.3]{Cormen2009}. Each
node~$x$ contains three fields: (\emph{i})~an interval
$x.\textit{in}$, represented by its lower and upper bounds;
(\emph{ii})~the minimum lower bound~$x.\textit{minlower}$ among all
intervals stored at the subtree rooted at~$x$; (\emph{iii})~the
maximum upper bound~$x.\textit{maxupper}$ among all intervals stored
at the subtree rooted at~$x$.

Insertions and deletions are handled according to the normal rules
for~AVL trees, with the additional requirement that any update of the
values of~$\textit{maxupper}$ and~$\textit{minlower}$ must be
propagated up to the root. During tree operations, nodes are kept
sorted according to the lower bounds. Since the height of an~AVL tree
is~$\Theta(\lg n)$, insertions and deletions in the augmented data
structure require~$O(\lg n)$ time in the worst case. Creating a new
tree with~$n$ nodes requires total time~$O(n \lg n)$ and space~$O(n)$.

\begin{algorithm}[t]
\caption{\textsc{Interval-Tree-Matching-1D}$(\mathbf{S}, \mathbf{U})$}\label{alg:itm}
\begin{algorithmic}[1]
\Function{Interval-Query}{$\textit{node}, q$}

\If{$\textit{node} = \textit{null}$ \OR\ $\textit{node}.\textit{maxupper} < q.\textit{lower}$ \OR\ $\textit{node}.\textit{minlower} > q.\textit{upper}$}
\State \textbf{return}
\EndIf

\State \Call{Interval-Query}{$\textit{node}.\textit{left}, q$}

\If{\Call{Intersect-1D}{$\textit{node}.\textit{in}, q$}}
\State \Call{Report}{$\textit{node}.\textit{in}, q$}
\EndIf

\If{$q.\textit{upper} \geq \textit{node}.\textit{in}.\textit{lower}$}
\State \Call{Interval-Query}{$\textit{node}.\textit{right}, q$}
\EndIf
\EndFunction

\Statex

\State $T \gets \textrm{create interval tree for $\mathbf{S}$}$\label{alg:int-tree-create}
\ForAll{update regions $u \in \mathbf{U}$ \textbf{in parallel}}\label{alg:int-tree-loop-begin}
\State \Call{Interval-Query}{$T.\textit{root}, u$}
\EndFor\label{alg:int-tree-loop-end}
\end{algorithmic}
\end{algorithm}

Algorithm~\ref{alg:itm} illustrates how
parallel~\ac{ITM} works. The first step is to build an interval
tree~$T$ containing the subscription intervals~$\mathbf{S}$ (the
details are omitted for the sake of conciseness;
Section~\ref{sec:experimental-evaluation} provides directions to the
source code). Function \Call{Interval-Query}{$T.\textit{root}, u$} is
then used to report all intersections among an update region~$u$ and
all subscription intervals stored in~$T$. The procedure is similar to
a conventional binary search tree lookup, using
the~$x.\textit{minlower}$ and~$x.\textit{maxupper}$ fields of every
node~$x$ to steer the visit away from irrelevant subtrees. Since each
node represents one interval, function \Call{Interval-Query}{} reports
each intersection only once.

\paragraph*{Asymptotic execution time}
An interval tree can be created in time~$O(n \lg n)$ like an
ordinary~AVL tree. To determine the cost of one invocation of function
\Call{Interval-Query}{} we first observe that each node can be visited
at most once per call, so~$O(n)$ is an upper bound on the asymptotic
running time of \Call{Interval-Query}{}. If region~$u$ overlaps
with~$K_u$ subscription intervals, then the execution time of
\Call{Interval-Query}{$T.\mathit{root}, u$} is also bound by~$O(K_u
\lg n)$; combining the two bounds we get that one call
costs~$O\left(\min\{n, K_u \lg n\}\right)$. Since
function~\Call{Interval-Query}{} is called~$m$ times (one for each
update region~$u$), the total query time is $O\left( \min\{m \times n,
K \lg n\}\right)$, $K$ being the number of intersections.

Once the tree is built, its structure is never modified. Therefore,
the iterations of the loop on
lines~\ref{alg:int-tree-loop-begin}--\ref{alg:int-tree-loop-end} can
be evenly split across~$P$ processors on a shared-memory architecture.
The parallel query time is then reduced by a factor~$P$ and becomes
$O\left( \min\{m \times n, K \lg n\} / P \right)$.

Observe that Algorithm~\ref{alg:itm} allows the roles
of~$\mathbf{S}$ and~$\mathbf{U}$ to be swapped. This can be helpful if
the number of update regions~$m$ is much lower than the number of
subscription regions~$m$. If $m \ll n$ it is more convenient to build
a tree on~$\mathbf{U}$ instead than on~$\mathbf{S}$, since the height
will be lower. With this optimization we can replace~$n$ with~$m$ in
the asymptotic query time above.

Different implementations of the interval tree are possible. Priority
search trees~\cite{McCreight85} support insertion and deletion in time
$O(\lg n)$, and can report all~$K$ intersections with a given query
interval in time~$O(K + \lg n)$. While the implementation above based
on~AVL trees is asymptotically less efficient, it has the advantage of
being easier to implement since it relies on a familiar data
structure. We have chosen~AVL trees over other balanced search trees,
such as red-black trees~\cite{rbtree}, because~AVL trees are more
rigidly balanced and therefore allow faster queries. It should be
observed that~\ac{ITM} is not tied to any specific interval tree
implementation; therefore, any data structure can be used as a drop-in
replacement.

\paragraph*{Dynamic interval management} 
An interesting feature of~\ac{ITM} is that it can easily cope with
\emph{dynamic} intervals. The~\ac{HLA} specification allows federates
to modify (i.e., move or grow/shrink) subscription and update regions;
the ability to do so is indeed essential in almost every agent-based
simulation model. A subscription region~$s \in \mathbf{S}$ (resp. $u
\in \mathbf{U}$) changing its position or size will trigger at
most~$O(m)$ (resp.~$O(n)$) new overlaps, so it makes sense to exploit
the data structures already built instead of running the matching
phase from scratch each time.

The interval tree data structure can be used to implement a dynamic
data distribution management scenario as follows.  We can use two
interval trees~$T_U$ and~$T_S$ holding the set of update and
subscription regions, respectively. If an update region~$u \in
\mathbf{U}$ is modified, we can identify the~$K_u$ subscriptions
overlapping with~$u$ in time $O\left( \min \{n, K_u \lg n \}\right)$
($O\left( \min \{n, K_u \lg n \} / P\right)$ if~$P$ processors are
used) by repeatedly calling the function \Call{Interval-Query}{}
on~$T_U$.  Similarly, if a subscription region~$s \in \mathbf{S}$
changes, the~$K_s$ overlaps with~$s$ can be computed in time $O\left(
\min\{m, K_s \lg m\}\right)$ using~$T_U$. When a subscription or
update region is modified, the appropriate tree must be updated by
deleting and re-inserting the node representing the region that
changed. These operations require time $O(\lg n)$ for the subscription
regions, and $O(\lg n)$ for the update regions.

%%%%%%%%%%%%%%%%%%%%%%%%%%%%%%%%%%%%%%%%%%%%%%%%%%%%%%%%%%%%%%%%%%%%%%%%%%%%%%
\section{Parallel Sort-based Matching}\label{sec:parallel-sort-matching}

The parallel~\ac{ITM} algorithm from the previous section relies on a
data structure (the interval tree) to efficiently enumerate all
intersections among a set of intervals with a given query
interval~$q$. Once built, the interval tree allows a high degree of
parallelism since all update regions can be compared concurrently with
the subscription regions stored in the tree. \ac{ITM} can be easily
extended to support dynamic management of subscription and update
regions.

We now propose a parallel solution to the region matching problem
based on a novel parallel algorithm derived from~\ac{SBM}. 
The parallel version of~\ac{SBM} will be derived incrementally,
starting from the serial version that has been described in
Section~\ref{sec:region-matching} (Algorithm~\ref{alg:sbm}). As we may
recall, \ac{SBM} operates in two phases: first, a list~$T$ of
endpoints of all regions is built and sorted; then, the sorted list is
traversed to compute the values of the~\texttt{SubSet}
and~\texttt{UpdSet} variables, from which the list of overlaps is
derived. Let us see if and how each step can be parallelized.

On a shared-memory architecture with~$P$ processors, building the list
of endpoints can be trivially parallelized, especially if this data
structure is realized with an array of $2 \times (n+m)$ elements
rather than a linked list. Sorting the endpoints can be realized using
a parallel sorting algorithm such as parallel mergesort~\cite{Cole88}
or parallel quicksort~\cite{Wheat92,Tsigas:2003}, both of which are
optimal. The traversal of the sorted list of endpoints
(Algorithm~\ref{alg:sbm}
lines~\ref{alg:sbm:loop-start}--\ref{alg:sbm:loop-end}) is, however,
more problematic. Ideally, we would like to evenly split the list~$T$
into~$P$ segments $T_0, \ldots, T_{P-1}$, and assign each segment to a
processor so that all segments can be processed
concurrently. Unfortunately, this is not possible due to the presence
of \emph{loop-carried dependencies}. A loop carried dependency is a
data dependence that causes the result of a loop iteration to depend
on previous iterations. In the~\ac{SBM} algorithm the loop-carried
dependencies are caused by the variables \texttt{SubSet} and
\texttt{UpdSet}, whose values depend on those computed on the previous
iteration.

\begin{algorithm}[t]
\caption{\textsc{Parallel-SBM-1D}$(\mathbf{S}, \mathbf{U})$}\label{alg:par-sbm}
\begin{algorithmic}[1]
\State $T \gets \emptyset$
\ForAll{regions $x \in \mathbf{S} \cup \mathbf{U}$ \textbf{in parallel}}
\State Insert $x.\textit{lower}$ and $x.\textit{upper}$ in $T$
\EndFor
\State Sort $T$ \textbf{in parallel}, in non-decreasing order
\State Split $T$ into $P$ segments $T_0, \ldots, T_{P-1}$
\State $\langle$Initialize $\texttt{SubSet}[0..P-1]$ and $\texttt{UpdSet}[0..P-1] \rangle$\label{alg:par-sbm-init}
\For{$p \gets 0~\textrm{to}~P-1$ \textbf{in parallel}}
\ForAll{endpoints $t \in T_p$ in non-decreasing order}
\If{$t$ belongs to subscription region $s$}
\If{$t$ is the lower bound of $s$}
\State $\texttt{SubSet}[p] \gets \texttt{SubSet}[p] \cup \{s\}$
\Else
\State $\texttt{SubSet}[p] \gets \texttt{SubSet}[p] \setminus \{s\}$
\ForAll{$u \in \texttt{UpdSet}[p]$} \Call{Report}{$s, u$}
\EndFor
\EndIf
\Else\Comment{$t$ belongs to update region $u$}
\If{$t$ is the lower bound of $u$}
\State $\texttt{UpdSet}[p] \gets \texttt{UpdSet}[p] \cup \{u\}$
\Else
\State $\texttt{UpdSet}[p] \gets \texttt{UpdSet}[p] \setminus \{u\}$
\ForAll{$s \in \texttt{SubSet}[p]$} \Call{Report}{$s, u$}
\EndFor
\EndIf
\EndIf
\EndFor
\EndFor
\end{algorithmic}
\end{algorithm}

Let us pretend that the scan phase \emph{could} be parallelized
somehow. Then, a parallel version of~\ac{SBM} would look like
Algorithm~\ref{alg:par-sbm} (line~\ref{alg:par-sbm-init} will be
explained shortly). The major difference between
Algorithm~\ref{alg:par-sbm} and its sequential counterpart is that the
former uses two arrays~$\texttt{SubSet}[p]$ and~$\texttt{UpdSet}[p]$
instead of the scalar variables~\texttt{SubSet}
and~\texttt{UpdSet}. This allows each processor to operate on its
private copy of the subscription and update sets, achieving the
maximum level of parallelism.

It is not difficult to see that Algorithm~\ref{alg:par-sbm} is
equivalent to sequential~\ac{SBM} (i.e., they produce the same result)
if and only if $\texttt{SubSet}[0..P-1]$ and $\texttt{UpdSet}[0..P-1]$
are properly initialized. Specifically, $\texttt{SubSet}[p]$ and
$\texttt{UpdSet}[p]$ must be initialized with the values that the
sequential~\ac{SBM} algorithm assigns to \texttt{SubSet} and
\texttt{UpdSet} right after the last endpoint of~$T_{p-1}$ is
processed, for every $p=1, \ldots, P-1$; $\texttt{SubSet}[0]$ and
$\texttt{UpdSet}[0]$ must be initialized to the empty set.

It turns out that $\texttt{SubSet}[0..P-1]$ and
$\texttt{UpdSet}[0..P-1]$ can be computed efficiently using a
\emph{parallel prefix computation} (also called \emph{parallel scan}
or \emph{parallel prefix-sum}). To make this paper self-contained, we
introduce the concept of prefix computation before illustrating the
missing part of the parallel~\ac{SBM} algorithm.

\paragraph*{Prefix computations}
A prefix computation consists of a sequence of~$N > 0$ data items
$x_0, \ldots, x_{N-1}$ and a binary associative
operator~$\oplus$. There are two types of prefix computations:
an \emph{inclusive scan} produces a new sequence of~$N$ data items $y_0,
\ldots, y_{N-1}$ defined as:

\begin{alignat*}{2}
y_0 &= x_0 \\
y_1 &= y_0 \oplus x_1 &&= x_0 \oplus x_1 \\
y_2 &= y_1 \oplus x_2 &&= x_0 \oplus x_1 \oplus x_2 \\
&\vdots \\
y_{N-1} &= y_{N-2} \oplus x_{N-1} &&= x_0 \oplus x_1 \oplus \ldots \oplus x_{N-1}
\end{alignat*}

\noindent while an \emph{exclusive scan} produces the sequence $z_0,
z_1, \ldots z_{N-1}$ defined as:

\begin{alignat*}{2}
z_0 &= 0 \\
z_1 &= z_0 \oplus x_0 &&= x_0 \\
z_2 &= z_1 \oplus x_1 &&= x_0 \oplus x_1 \\
&\vdots \\
z_{N-1} &=  z_{N-2} \oplus x_{N-2} &&= x_0 \oplus x_1 \oplus \ldots \oplus x_{N-2}
\end{alignat*}

\noindent where~$0$ is the neutral element of operator $\oplus$, i.e.,
$0 \oplus x = x$.

Hillis and Steele~\cite{Hillis:1986} proposed a parallel algorithm for
computing the prefix sum of~$N$ items with~$N$ processors in~$O(\lg
N)$ parallel steps and total work~$O(N \lg N)$. This result was
improved by Blelloch~\cite{Blelloch89} who described a parallel
implementation of the scan primitive of~$N$ items on~$P$ processors
requiring time $O(N/P + \lg P)$. Blelloch's algorithm is optimal
when~$N > P \lg P$, meaning that the total amount of work it performs
over all processors is the same as the (optimal) serial algorithm for
computing prefix sums, i.e., $O(N)$.

\begin{figure*}[t]
\centering\includegraphics[scale=.8]{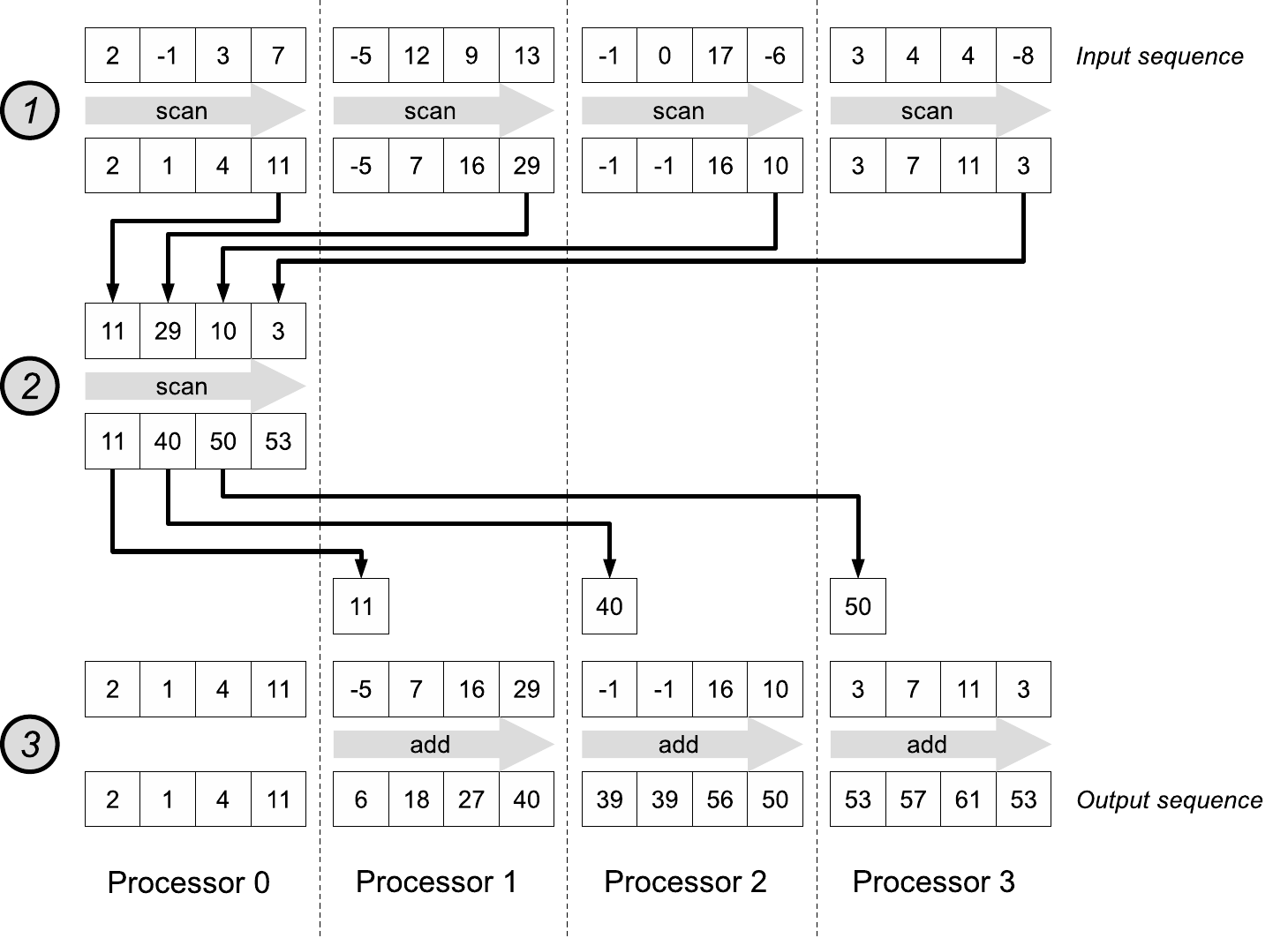}
\caption{Parallel prefix sum computation.}\label{fig:scan}
\end{figure*}

A somewhat simpler algorithm for computing prefix sums of~$N$ items
with~$P$ processors in time $O(N/P + P)$ is illustrated in
Figure~\ref{fig:scan}; in the figure we consider the addition
operator, although the idea applies to any associative
operator~$\oplus$. The computation involves two parallel steps (steps
\circlenum{1} and \circlenum{3} in the figure), and one serial step
(step \circlenum{2}). In step~\circlenum{1} the input sequence is
split across the processors, and each processor computes the prefix
sum of the elements in its portion.  In step~\circlenum{2} the master
computes the prefix sum of the~$P$ last local sums.  Finally, in
step~\circlenum{3} the master scatters the first~$(P-1)$ computed
values (prefix sums of the last local sums) to the last~$(P-1)$
processors. Each processor, except the first one, adds (more
precisely, applies the~$\oplus$ operator) the received value to the
prefix sums from step~\circlenum{1}, producing a portion of the output
sequence.

Steps~\circlenum{1} and~\circlenum{3} require time~$O(N/P)$ each,
while step~\circlenum{2} is executed by the master in time $O(P)$,
yielding a total time~$O(N/P + P)$. Therefore, the algorithm is
optimal when~$N > P^2$. Since the current generation of CPUs have a
small number of cores (e.g., $P \leq 72$ for the Intel Xeon Phi) and
the number of regions~$N$ is usually very large, the algorithm above
can be considered optimal for any practical purpose. We remark that
the parallel~\ac{SBM} algorithm can be readily implemented with the
tree-structured reduction operation, and therefore will still be
competitive on future generations of processors with a higher number
of cores.

\begin{algorithm}[h]
\caption{$\langle$Initialize $\texttt{SubSet}[0..P\!-\!1]$ and $\texttt{UpdSet}[0..P\!-\!1] \rangle$}\label{alg:par-sbm-cont}
\begin{algorithmic}[1]
\For{$p \gets 0~\textrm{to}~P-1$ \textbf{in parallel}}\label{alg:par-sbm-loop-begin}\Comment{Executed by all processors in parallel}
\State $\texttt{Sadd}[p] \gets \emptyset$, $\texttt{Sdel}[p] \gets \emptyset$, $\texttt{Uadd}[p] \gets \emptyset$, $\texttt{Udel}[p] \gets \emptyset$
\ForAll{points $t \in T_p$ in non-decreasing order}
\If{$t$ belongs to subscription region $s$}
\If{$t$ is the lower bound of $S_i$}
\State $\texttt{Sadd}[p] \gets \texttt{Sadd}[p] \cup \{s\}$
\ElsIf{$s \in \texttt{Sadd}[p]$}
\State $\texttt{Sadd}[p] \gets \texttt{Sadd}[p] \setminus \{s\}$
\Else
\State $\texttt{Sdel}[p] \gets \texttt{Sdel}[p] \cup \{s\}$
\EndIf
\Else\Comment{$t$ belongs to update region $u$}
\If{$t$ is the lower bound of $u$}
\State $\texttt{Uadd}[p] \gets \texttt{Uadd}[p] \cup \{u\}$
\ElsIf{$u \in \texttt{Uadd}[p]$}
\State $\texttt{Uadd}[p] \gets \texttt{Uadd}[p] \setminus \{u\}$
\Else
\State $\texttt{Udel}[p] \gets \texttt{Udel}[p] \cup \{u\}$
\EndIf
\EndIf
\EndFor
\EndFor\label{alg:par-sbm-loop-end}
\Statex\Comment{Executed by the master only}
\State $\texttt{SubSet}[0] \gets \emptyset$, $\texttt{UpdSet}[0] \gets \emptyset$\label{alg:par-sbm-step-2-begin}
\For{$p \gets 1~\textrm{to}~P-1$}
\State $\texttt{SubSet}[p] \gets \texttt{SubSet}[p-1] \cup \texttt{Sadd}[p-1] \setminus \texttt{Sdel}[p-1]$
\State $\texttt{UpdSet}[p] \gets \texttt{UpdSet}[p-1] \cup \texttt{Uadd}[p-1] \setminus \texttt{Udel}[p-1]$
\EndFor\label{alg:par-sbm-step-2-end}
\end{algorithmic}
\end{algorithm}

\begin{figure*}[t]
\centering\includegraphics[scale=.8]{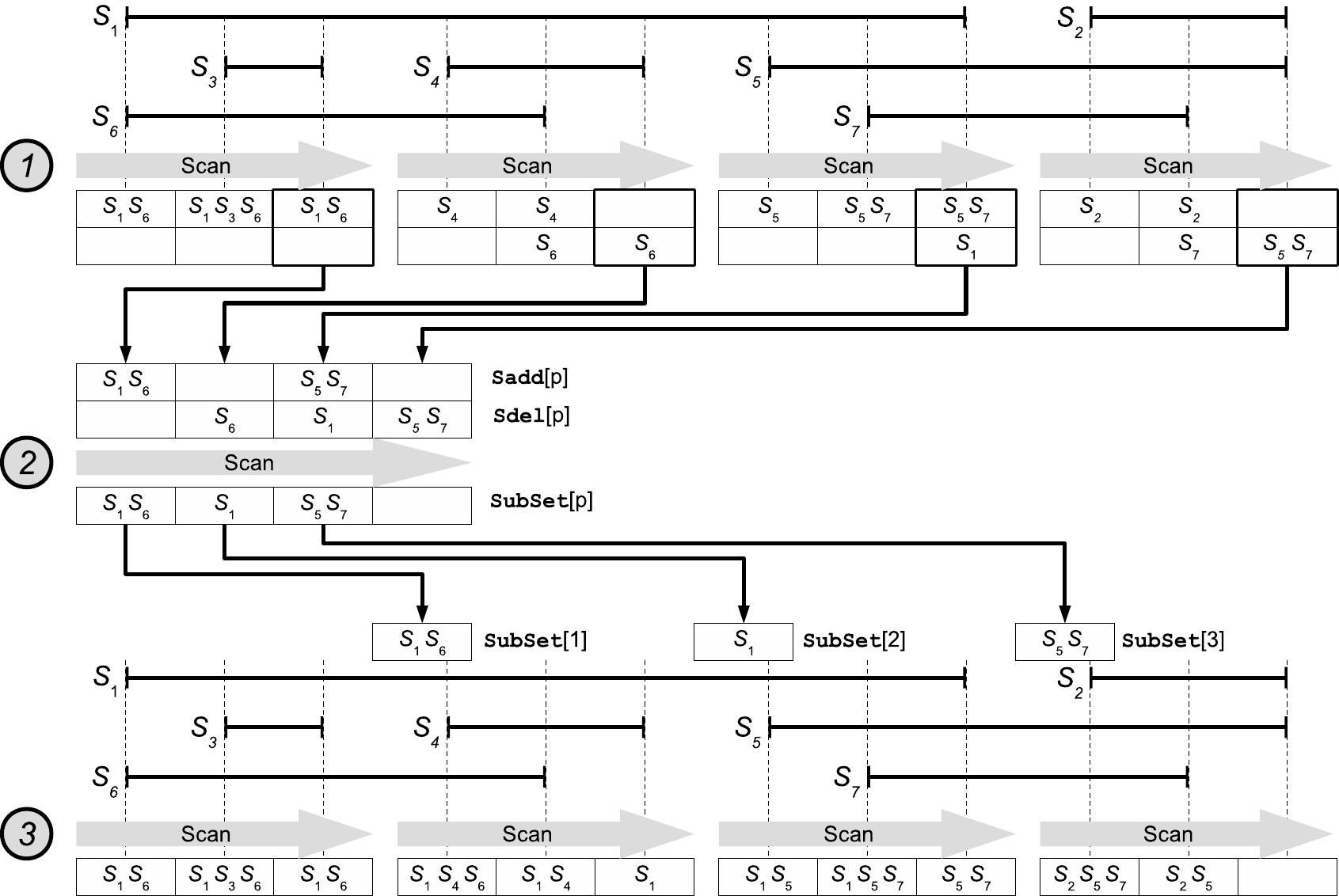}
\caption{Parallel prefix computation for the~\ac{SBM} algorithm.}\label{fig:par-sbm}
\end{figure*}

\paragraph*{Initialization with prefix computation}
We can now complete the description of the parallel~\ac{SBM} algorithm
by showing how the arrays $\texttt{SubSet}[p]$ and
$\texttt{UpdSet}[p]$ can be initialized in parallel. To better
illustrate the steps involved, we refer to the example in
Figure~\ref{fig:par-sbm}; in the figure we consider subscription
regions only, since the procedure for update regions is the same.

The sorted array of endpoints~$T$ is evenly split into~$P$ segments
$T_0, \ldots, T_{P-1}$ of~$2\times(n+m)/P$ elements each. Processor~$p$
scans the endpoints $t \in T_p$ in non-decreasing order, updating four
auxiliary variables $\texttt{Sadd}[p]$, $\texttt{Sdel}[p]$,
$\texttt{Uadd}[p]$, and $\texttt{Udel}[p]$. Informally,
$\texttt{Sadd}[p]$ and $\texttt{Sdel}[p]$ (resp. $\texttt{Uadd}[p]$
and $\texttt{Udel}[p]$) contain the endpoints that the
sequential~\ac{SBM} algorithm would add/remove from \texttt{SubSet}
(resp. \texttt{UpdSet}) while scanning the endpoints belonging to
segment $T_p$. More formally, at the end of each local scan the
following invariants hold:
\begin{enumerate}
\item $\texttt{Sadd}[p]$ (resp. $\texttt{Uadd}[p]$) contains the
  subscription (resp. update) intervals whose lower endpoint belongs
  to $T_p$, and whose upper endpoint does not belong to $T_p$;
\item $\texttt{Sdel}[p]$ (resp. $\texttt{Udel}[p]$) contains the
  subscription (resp. update) intervals whose upper endpoint belongs
  to $T_p$, and whose lower endpoint does not belong to $T_p$.
\end{enumerate}
This step is realized by
lines~\ref{alg:par-sbm-loop-begin}--\ref{alg:par-sbm-loop-end} of
Algorithm~\ref{alg:par-sbm-cont}, and its effects are shown in
Figure~\ref{fig:par-sbm}~\circlenum{1}. The figure reports the values
of $\texttt{Sadd}[p]$ and $\texttt{Sdel}[p]$ after each endpoint has
been processed; the algorithm does not store every intermediate value,
since only the last ones (within thick boxes) will be needed by the
next step.

Once all $\texttt{Sadd}[p]$ and $\texttt{Sdel}[p]$ are available, the
next step is executed by the master and consists of computing the
values of $\texttt{SubSet}[p]$ and $\texttt{UpdSet}[p]$, $p=0, \ldots,
P-1$. Recall from the discussion above that $\texttt{SubSet}[p]$
(resp. $\texttt{UpdSet}[p]$) is the set of active subscription
(resp. update) intervals that would be identified by the
sequential~\ac{SBM} algorithm right after the end of segment $T_0 \cup
\ldots \cup T_{p-1}$. The values of $\texttt{SubSet}[p]$ and
$\texttt{UpdSet}[p]$ are related to $\texttt{Sadd}[p]$,
$\texttt{Sdel}[p]$, $\texttt{Uadd}[p]$ and $\texttt{Udel}[p]$ as
follows:
\begin{align*}
\texttt{SubSet}[p] &= \begin{cases}
\emptyset & \mbox{if $p=0$} \\
\texttt{SubSet}[p-1] \cup \texttt{Sadd}[p-1] \setminus \texttt{Sdel}[p-1] & \mbox{if $p>0$}
\end{cases} \\
\texttt{UpdSet}[p] &= \begin{cases}
\emptyset & \mbox{if $p=0$} \\
\texttt{UpdSet}[p-1] \cup \texttt{Uadd}[p-1] \setminus \texttt{Udel}[p-1] & \mbox{if $p>0$}
\end{cases}
\end{align*}
Intuitively, the set of active intervals at the end of~$T_p$ can be
computed from those active at the end of~$T_{p-1}$, plus the intervals
that became active in~$T_p$, minus those that ceased to be active
in~$T_p$.

Lines~\ref{alg:par-sbm-step-2-begin}--\ref{alg:par-sbm-step-2-end} of
Algorithm~\ref{alg:par-sbm-cont} take care of this computation; see
also Figure~\ref{fig:par-sbm}~\circlenum{2} for an example. Once the
initial values of $\texttt{SubSet}[p]$ and $\texttt{UpdSet}[p]$ have
been computed, Algorithm~\ref{alg:par-sbm} can be resumed to identify
the list of overlaps.

\paragraph*{Asymptotic execution time} 
We now analyze the asymptotic execution time of parallel \ac{SBM}.
Let~$N$ denote the total number of subscription and update regions,
and~$P$ the number of processors.  Algorithm~\ref{alg:par-sbm}
consists of three phases:
\begin{enumerate}
\item Sorting~$T$ in non-decreasing order requires total time $O\left(
  (N \lg N) / P\right)$ using a parallel sorting algorithm such as
  parallel mergesort~\cite{Cole88}.
\item Computing the initial values of $\texttt{SubSet}[p]$ and
  $\texttt{UpdSet}[p]$ for each $p=0, \ldots, P-1$ requires
  $O\left(N/P + P\right)$ steps using the two-level scan shown on
  Algorithm~\ref{alg:par-sbm-cont}; the time could be reduced to
  $O\left(N/P + \lg P\right)$ steps using the tree-structured scan by
  Blelloch~\cite{Blelloch89}.
\item Each of the final local scans require $O(N/P)$ steps.
\end{enumerate}

Note, however, that phases~2 and~3 require the manipulation of data
structures to store sets of endpoints, supporting insertions and
removals of single elements and whole sets. Therefore, a single step
of the algorithm has a non-constant time complexity that depends on
the actual implementation of sets and the number of elements they
contain; this issue will be discussed in more detail in
Section~\ref{sec:experimental-evaluation}. During phase~3 total
time~$O(K)$ is spent cumulatively by all processors to report all~$K$
intersections.

\paragraph*{Some remarks on distributed-memory and GPU implementations}
Although the focus of this paper is on shared-memory architectures, we
provide here some remarks on possible distributed-memory and~\ac{GPU}
implementations of Algorithm~\ref{alg:par-sbm}.
  
In a distributed-memory system, computing nodes exchange information
through some type of high-performance network connection. It turns out
that a distributed-memory implementation of
Algorithm~\ref{alg:par-sbm} can be realized with minimal
modifications. First, a suitable distributed-memory sorting algorithm
(e.g., \cite{parallel-sorting}) can be used to sort the list of
endpoints. Then, the parallel prefix computation shown in
Figure~\ref{fig:par-sbm} can be realized efficiently since it is based
on the \emph{Scatter/Gather} communication
pattern~\cite{Mattson:2004}. A \emph{Scatter} operation allows a
single process to send portions of a local array to multiple
destinations, and is executed between steps~\circlenum{2}
and~\circlenum{3} in Figure~\ref{fig:par-sbm}. The symmetric
\emph{Gather} allows multiple processes to send portions of an array
to a single destination where they are concatenated; this is required
between steps~\circlenum{1} and~\circlenum{2}. Since Scatter/Gather
primitives are very useful in many contexts, they are efficiently
supported by software middlewares (e.g., the \texttt{MPI\_Scatter()}
\texttt{MPI\_Gather()} functions of the Message Passing Interface
specification), or directly at the hardware level~\cite{Almasi:2005}.

An efficient~\ac{GPU} implementation of Algorithm~\ref{alg:par-sbm},
however, poses several challenges. Although GPU-based efficient
algorithms for sorting and doing prefix computations are
available~\cite{Thrust}, the data structure used to represent sets of
endpoints must be designed carefully. As described earlier in this
Section, Algorithm~\ref{alg:par-sbm} performs~$\Theta(N)$ set
operations (unions and differences) during the prefix computation, and
therefore the data structure used for representing sets must be chosen
wisely (we will return to this issue in
Section~\ref{sec:experimental-evaluation}). Data structures based on
hash tables or trees are problematic on the~\ac{GPU}, although not
impossible~\cite{Awad:2019, GPU-hash-tables}. A simpler implementation
of sets using bit vectors appears to be better suited: a bit vector is
a sequence of~$N$ bits, where item~$i$ is in the set if and only if
the $i$-th bit is one. Bit vectors allow union, intersection and set
difference to be realized efficiently using nothing more than Boolean
operators; however, bit vectors require considerable amounts of memory
if the number~$N$ of items that could be in the set is large. This
issue requires further investigation, and is subject of ongoing
research.

%%%%%%%%%%%%%%%%%%%%%%%%%%%%%%%%%%%%%%%%%%%%%%%%%%%%%%%%%%%%%%%%%%%%%%%%%%%%%%
\section{Experimental Evaluation}\label{sec:experimental-evaluation}

In this section we evaluate the performance and scalability of the
parallel~\ac{ITM} and parallel~\ac{SBM} algorithms, and compare them
to parallel implementations of~\ac{BFM} and~\ac{GBM}. \ac{BFM}
and~\ac{ITM} have been implemented in C, while~\ac{SBM} and~\ac{GBM}
have been implemented in C++. To foster the reproducibility of our experiments, 
all the source code used in this performance evaluation is freely available on the research group
website~\cite{pads} with a Free Software license.

We used the GNU C Compiler (GCC)
version~4.8.4 with the \verb+-O3 -fopenmp -D_GLIBCXX_PARALLEL+ flags
to turn on optimization and to enable parallel constructs at the
compiler and library levels. Specifically, the \verb+-fopenmp+ flag
instructs the compiler to handle OpenMP directives in the source
code~\cite{OpenMP}, while the flag \verb+-D_GLIBCXX_PARALLEL+ enables parallel
implementations of some algorithms from the 
%C++~\ac{STL} 
C++ Standard Template Library (STL)
to be used
instead of their sequential counterparts (more details below).

OpenMP is an open interface supporting shared memory parallelism in
the C, C++ and FORTRAN programming languages. OpenMP allows the
programmer to label specific sections of the source code as parallel
regions; the compiler takes care of dispatching these regions to
different threads that can be executed by the available processors or
cores. In the C/C++ languages, OpenMP directives are specified using
\verb+#pragma+ directives. The OpenMP standard also defines some
library functions that can be called by the developer to query and
control the execution environment programmatically.

The~\ac{BFM} and~\ac{ITM} algorithms are \emph{embarrassingly
  parallel}, meaning that the iterations on their main loop
(line~\ref{alg:bfm:for} on Algorithm~\ref{alg:bfm} and
line~\ref{alg:int-tree-loop-begin} on Algorithm~\ref{alg:itm}) are
free of data races. Therefore, a single
\verb+#pragma omp parallel for+ directive is sufficient to distribute
the iterations of the loops across the processors.

\ac{GBM} requires more care, as we have already observed in
Section~\ref{sec:region-matching}, since the first loop
(lines~\ref{alg:gbm:first-begin}--\ref{alg:gbm:first-end},
Algorithm~\ref{alg:gbm}) has a data race due to possible concurrent
updates to the list~$G[i]$ in line~\ref{alg:gbm:list-add}. A simple
solution consists on protecting the statement with a
\verb+#pragma omp critical+ directive that ensures that only one
thread at a time can modify a list. However, this might limit the
degree of parallelism since two OpenMP threads would be prevented from
updating two different lists concurrently. To investigate this issue
we developed an ad-hoc, lock-free linked list data structure that
supports concurrent append operations without the need to declare a
critical section. Measurements showed that in our experiments the
ad-hoc linked list did not perform significantly better, so we decided
to use the standard \verb+std::list+ container provided by the
C++~\ac{STL} library~\cite{Stroustrup13}, and protect concurrent
updates with the OpenMP \verb+critical+ directive.

Our implementation of parallel~\ac{SBM} relies on some of the data
structures and algorithms provided by the C++~\ac{STL}. Specifically,
to sort the endpoints we use the parallel \verb+std::sort+ function
provided by the~\ac{STL} extensions for
parallelism~\cite{parallel-stl}. Indeed, the GNU~\ac{STL} provides
several parallel sort algorithms (multiway mergesort and quicksort
with various splitting heuristics) that are automatically selected at
compile time when the \verb+-D_GLIBCXX_PARALLEL+ compiler flag is
used. The rest of the~\ac{SBM} algorithm has been parallelized using
explicit OpenMP \verb+parallel+ directives.

The~\acf{SBM} algorithm relies on a suitable data structure to store
the sets of endpoints \texttt{SubSet} and \texttt{UpdSet} (see
Algorithms~\ref{alg:par-sbm} and~\ref{alg:par-sbm-cont}).
Parallel~\ac{SBM} puts a higher strain on this data structure than its
sequential counterpart, since it also requires efficient support for
unions and differences between sets, in addition to insertions and
deletions of single elements. We have experimented with several
implementations for sets: (\emph{i})~bit vectors based on the
\verb+std::vector<bool>+ \ac{STL} container; (\emph{ii})~an ad-hoc
implementation of bit vectors based on raw memory manipulation;
(\emph{iii})~the \verb+std::set+ container, that in the case of the
GNU~\ac{STL} is based on Red-Black trees~\cite{Bayer1972};
(\emph{iv})~the \verb+std::unordered_set+ container from the~2011
ISO~C++ standard, that is usually implemented using hash tables;
(\emph{v})~the \texttt{boost::dynamic\_bitset} container provided by
the Boost C++ library~\cite{boost}. The most efficient turned out to
be the \verb+std::set+ container, and it has been used in the
experiments described below.

\begin{table}[t]
  \centering\begin{tabular}{lll}
  \toprule
CPU              & Intel Xeon E5-2640 \\
Clock frequency  & 2.00 GHz           \\
Processors       & 2                  \\
Cores/proc       & 8                  \\
Total cores      & 16                 \\
Threads/core     & 2                  \\
HyperThreading   & Yes                \\
RAM              & 128 GB             \\
L3 cache size    & 20480 KB           \\
Operating System & Ubuntu~16.04.3~LTS \\
\bottomrule
\end{tabular}
\caption{Hardware used for the experimental
  evaluation.}\label{tab:spec}
\end{table}

\paragraph*{Experimental setup}
The experiments below have been carried out on a dual-socket,
octa-core server whose hardware specifications are shown in
Table~\ref{tab:spec}. The processors employ the~\ac{HT}
technology~\cite{HT}: in \ac{HT}-enabled CPUs some functional
components are duplicated, but there is a single main execution unit
for physical core. From the point of view of the~\ac{OS}, \ac{HT}
provides two logical processors for each physical core. Studies from
Intel and others have shown that in typical applications \ac{HT}
contributes a performance boost in the range
$16$--$28\%$~\cite{HT}. When two processes are executed on the same
core, they compete for the shared hardware resources resulting is
lower efficiency than the same two processes executed on two different
physical cores.

The number~$P$ of OpenMP threads to use can be chosen either
programmatically through the appropriate OpenMP functions, or setting
the \verb+OMP_NUM_THREADS+ environment variable. In our experiments,
$P$ never exceeds the total number of (logical) cores, so that
over-provisioning never happens. By default, the Linux scheduler 
spreads processes to different physical cores as long as
possible; only when there are more runnable processes than physical
cores does~\ac{HT} come into effect. All tests have been executed with
this default behavior.

For better comparability of our results with those in the literature,
we consider~$d=1$ dimensions and use the methodology and parameters
described in~\cite{Raczy2005} (a performance evaluation based on real
dataset will be described at the end of this section). The first
parameter is the total number of regions~$N$, that includes~$n=N/2$
subscription and~$m=N/2$ update regions. All regions have the same
length~$l$ and are randomly placed on a segment of total
length~$L=10^6$. $l$ is defined in such a way that a given overlapping
degree~$\alpha$ is obtained, where

\begin{equation*}
\alpha=\frac{\sum \text{area of regions}}{\text{area of the routing space}} = \frac{N \times l}{L}
\end{equation*}

Therefore, given~$\alpha$ and~$N$, the value of~$l$ is set to $l =
\alpha L / N$. The overlapping degree is an indirect measure of the
total number of intersections among subscription and update
regions. While the performance of~\ac{BFM} and~\ac{SBM} is not
affected by the number of intersections, this is not the case
for~\ac{ITM}, as will be shown below. We considered the same values
for~$\alpha$ as in~\cite{Raczy2005}, namely~$\alpha \in \{0.01, 1,
100\}$.  Finally, each measure is the average of~$50$ independent runs
to get statistically valid results. Our implementations do not
explicitly store the list of intersections, but only count them. We
did so to ensure that the execution times are not affected by the
choice of the data structure used to store the list of
intersections.

% PB vs. pITM vs. pSBM with 1M, alfa=100
\begin{figure*}[t]
  \centering%
  \subfigure[Wall-clock time\label{fig:wct-1M}]{\includegraphics[width=.45\textwidth]{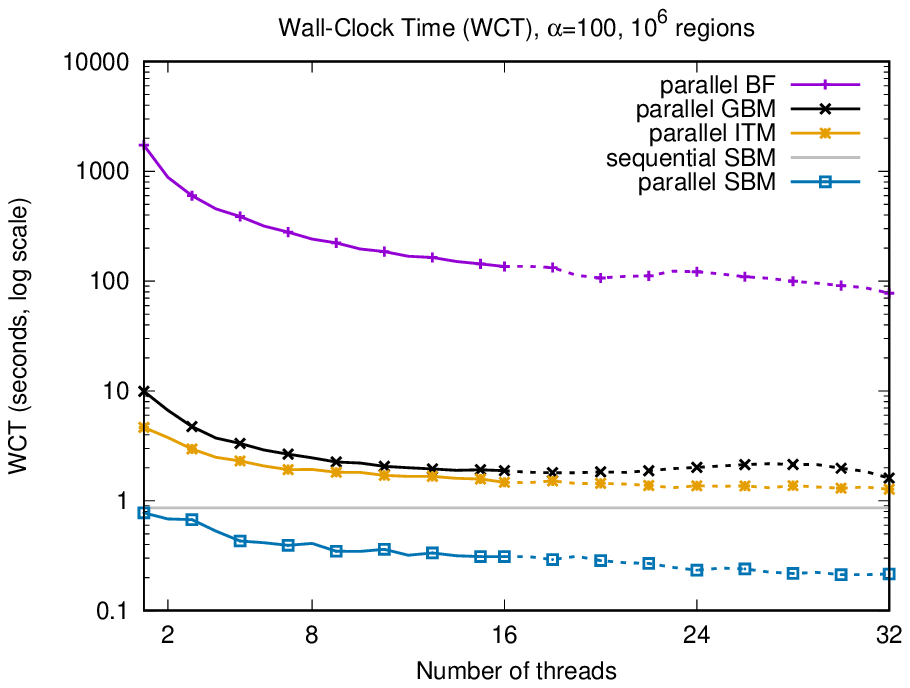}}\quad
  \subfigure[Speedup\label{fig:speedup-1M}]{\includegraphics[width=.45\textwidth]{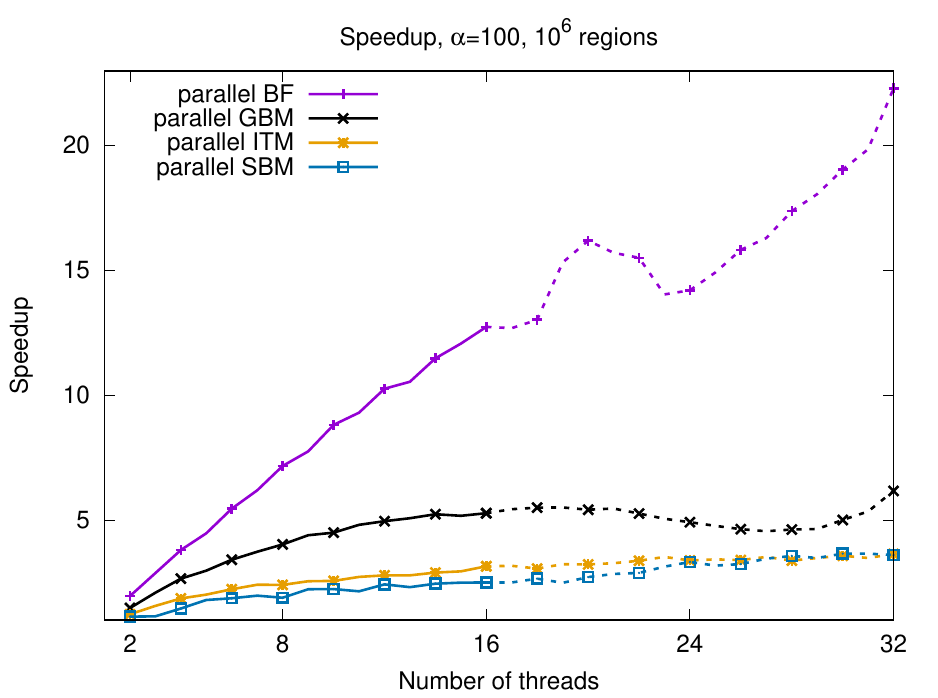}}
  \caption{Wall clock time and speedup of
    parallel~\{\ac{BFM},~\ac{GBM}, \ac{ITM},~\ac{SBM}\} with~$N=10^6$
    regions and overlapping degree~$\alpha=100$; the~\ac{GBM}
    algorithm uses~$3000$ regions. Dashed lines indicate the region
    where the number of OpenMP threads exceeds the number of
    (physical) CPU cores.}\label{fig:wct-speedup-1M}
\end{figure*}

\paragraph*{Wall clock time and Speedup}
The first metric we analyze is the~\ac{WCT} of the parallel
programs. The~\ac{WCT} includes the time needed to initialize all
ancillary data structures used by each algorithm (e.g., the time
needed to build the interval tree, or to fill the grid cells), but
does not include the time required to randomly initialize the input
regions.

Figure~\ref{fig:wct-1M} shows the \ac{WCT} for the parallel versions
of \ac{BFM}, \ac{GBM}, \ac{ITM} and \ac{SBM} as a function of the
number~$P$ of OpenMP threads, given~$N=10^6$ regions and overlapping
degree~$\alpha = 100$.  The~\ac{GBM} algorithm requires the user to
define the number of grid blocks ($\mathit{nblocks}$) to use. This
parameter should be carefully chosen since it affects the algorithm's
performance~\cite{Tan:2000}.  We have empirically determined that the
best running time with $P=32$ OpenMP threads with the parameters above
is~$3000$ regions. Dashed lines indicate when~$P$ exceeds the number
of physical cores.

We observe that the parallel~\ac{BFM} algorithm is about three orders
of magnitude slower than parallel~\ac{SBM}. This is unsurprising,
since the computational cost of~\ac{BFM} grows quadratically with the
number of regions (see Section~\ref{sec:region-matching}), while that
of~\ac{SBM} and~\ac{ITM} grows only polylogarithmically. \ac{SBM}
performs better than~\ac{BFM} by almost two orders of
magnitude. \ac{ITM} is faster than~\ac{BFM} (remember that
Figure~\ref{fig:wct-1M} uses a logarithmic scale), and provides the
additional advantage of requiring no tuning of parameters.

A drawback of~\ac{GBM} is that it requires the number of grid cells to
be defined. The optimal value depends both on the simulation model and
also on the number of OpenMP threads~$P$; our chosen value ($3000$
regions) is optimal for $P=32$, but not necessarily for the other
values of~$P$; indeed, we observe that the execution time of
parallel~\ac{GBM} increases around $P=24$; this shows up as a
prominent feature in the speedup graph as explained below.

The \emph{relative speedup} measures the increase in speed that a
parallel program achieves when more processors are employed to solve a
problem of the same size. This metric can be computed from
the~\ac{WCT} as follows. Let~$T(N, P)$ be the \ac{WCT} required to
process an input of size~$N$ using~$P$ processes (OpenMP
threads). Then, for a given~$N$, the relative speedup~$S_N(P)$ is
defined as $S_N(P) = T(N, 1) / T(N, P)$. Ideally, the maximum value of
$S_N(P)$ is $P$, which means that solving a problem with~$P$
processors requires~$1/P$ the time needed by a single processor. In
practice, however, several factors limit the speedup, such as the
presence of serial regions in the parallel program, uneven load
distribution, scheduling overhead, and heterogeneity in the execution
hardware. 

Figure~\ref{fig:speedup-1M} shows the speedups of the parallel
versions of~\ac{BFM}, \ac{GBM}, \ac{ITM} and~\ac{SBM} as a function of
the number of OpenMP threads~$P$; the speedup has been computed using
the wall clock times of Figure~\ref{fig:wct-1M}. Again, dashed lines
indicate data points where~$P$ exceeds the number of physical
processor cores. The~\ac{BFM} algorithm, despite being the less
efficient, is the most scalable due to its embarrassingly parallel
structure and lack of any serial part. \ac{SBM}, on the other hand, is
the most efficient but less scalable. \ac{SBM} achieves a~$2.6\times$
speedup with~$16$ OpenMP threads. When all ``virtual'' cores are used,
the speedup grows to $3.6\times$. The limited scalability of~\ac{SBM}
is somewhat to be expected, since its running time is very small and
therefore the overhead introduced by~OpenMP becomes non-negligible.

The effect of~\ac{HT} (dashed lines) is clearly visible in
Figure~\ref{fig:speedup-1M}. The speedup degrades when~$P$ exceeds the
number of cores, as can be seen from the different slopes for~\ac{BFM}
on \texttt{titan}. When \ac{HT} kicks in, load unbalance arises due to
contention of the shared control units of the processor cores, and
this limits the scalability.  The curious ``bulge'' that appears on
the curve is due to some OpenMP scheduling issues on the machine used
for the test, that is based on a~\ac{NUMA}
architecture~\cite{Broquedis2008}. Indeed, the bulge appears even if
we replace the body of the inner loop of the~\ac{BFM} algorithm
(line~\ref{alg:bfm:body} of algorithm~\ref{alg:bfm}) with a dummy
statement; moreover, the bulge does \emph{not} appear if we run
the~\ac{BFM} algorithm on a non-\acs{NUMA} machine.

% pITM vs. pSBM with 100M, alfa=100
\begin{figure}[ht]
  \centering%
  \subfigure[Wall-clock time\label{fig:wct-100M}]{\includegraphics[width=.45\textwidth]{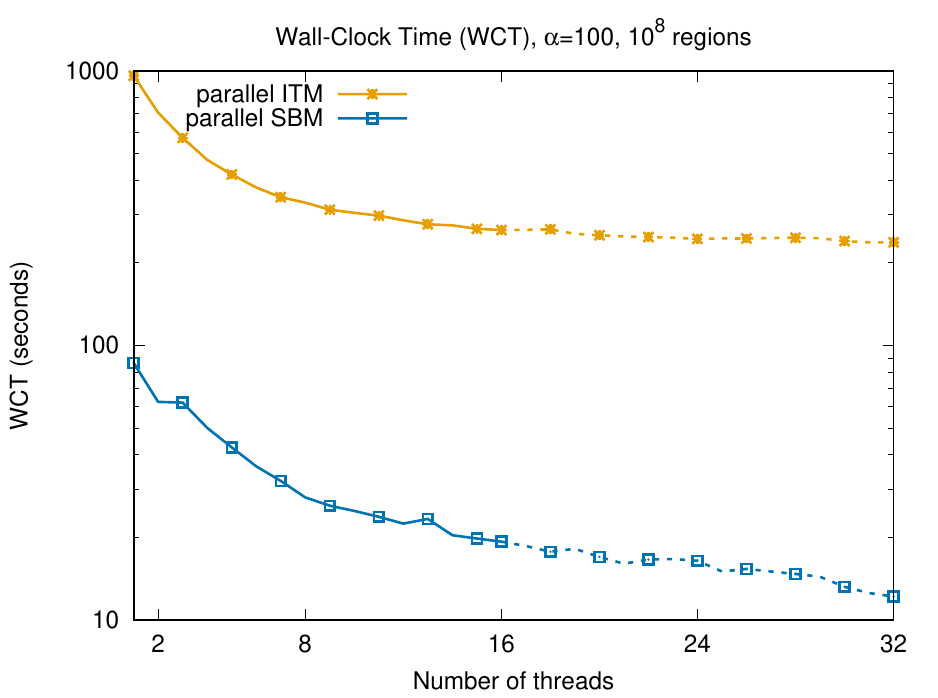}}\quad
  \subfigure[Speedup\label{fig:speedup-100M}]{\includegraphics[width=.45\textwidth]{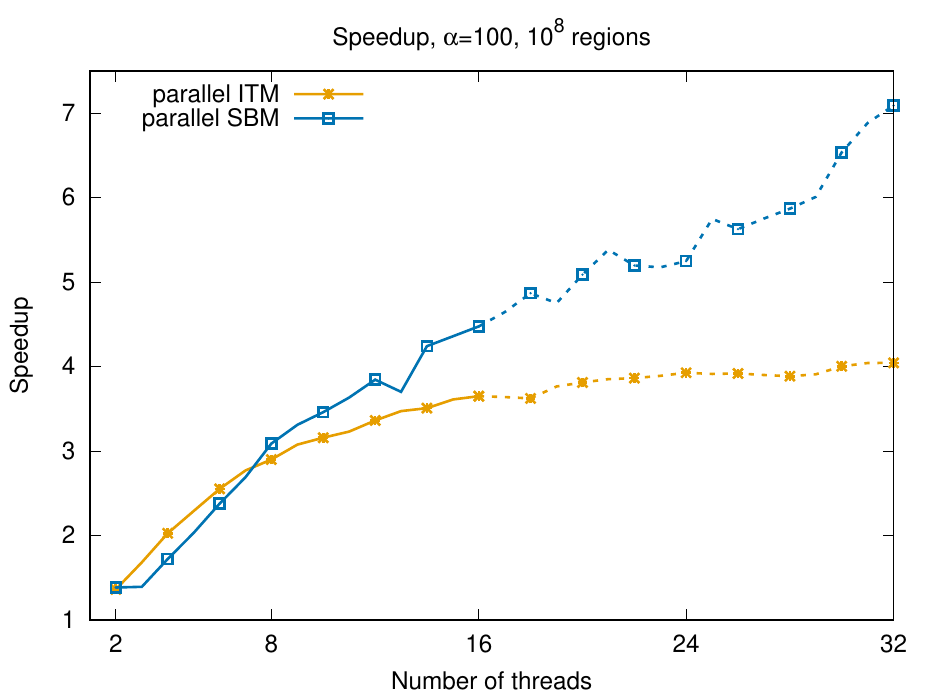}}
  \caption{Wall clock time and speedup of
    parallel~\{\ac{ITM},~\ac{SBM}\} with~$N=10^8$
    regions and overlapping degree~$\alpha=100$. Dashed lines indicate
    the region where the number of OpenMP threads exceeds the number
    of CPU cores, and therefore \ac{HT} comes into play.}\label{fig:wct-speedup-100M}
\end{figure}

The speedup of~\ac{SBM} improves if we increase the work performed by
the algorithm. Figure~\ref{fig:speedup-100M} shows the speedup of
parallel~\ac{ITM} and~\ac{SBM} with~$N=10^8$ regions and overlapping
degree~$\alpha = 100$; in this scenario both~\ac{BFM} and~\ac{GBM}
take so long that they have been omitted. The~\ac{SBM} algorithm
behaves better, achieving a $7\times$ speedup with $P=32$ threads.
The reason of this improvement is that increasing the amount of work
executed by each processor reduces the synchronization overhead, which
is particularly beneficial on multi-socket \ac{NUMA} machines.

\begin{figure}[t]
\centerline{\includegraphics[scale=.7]{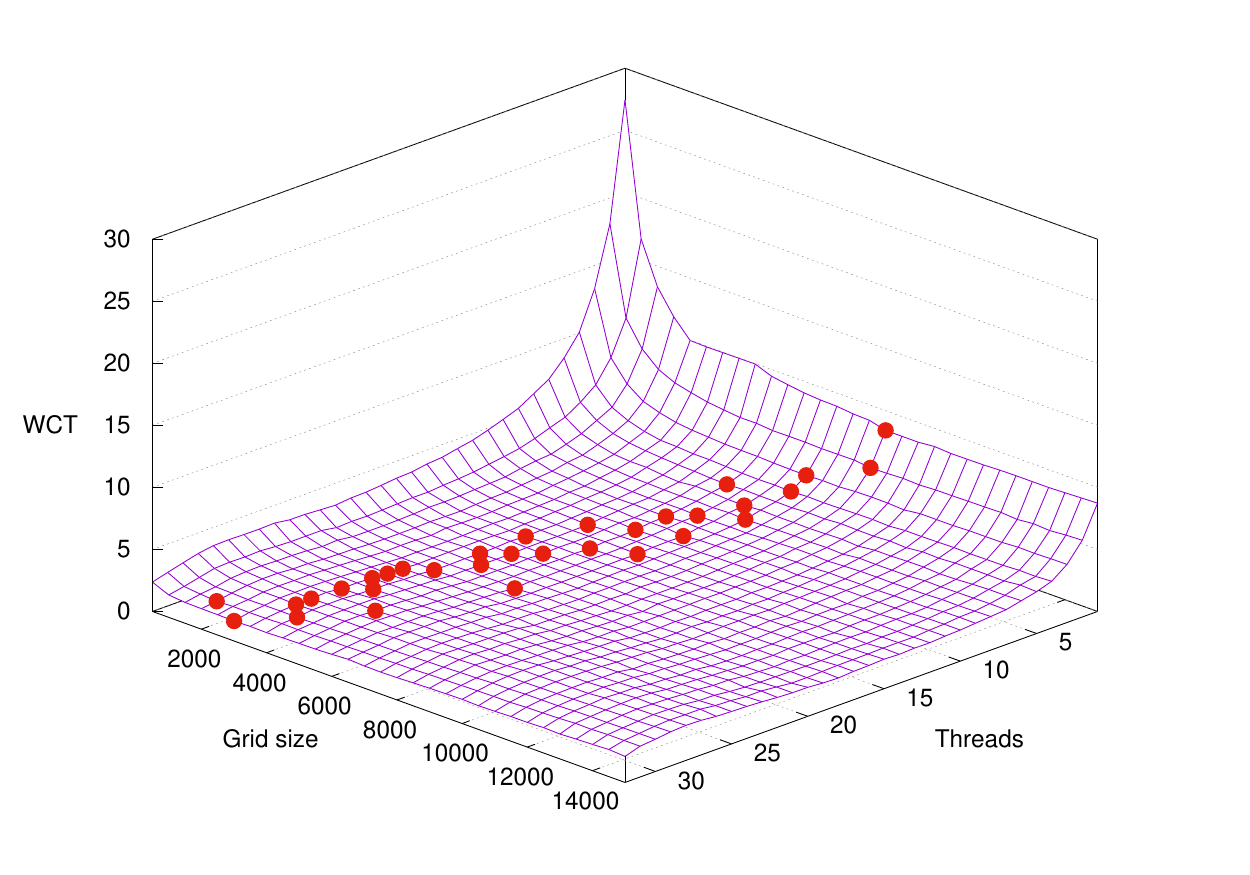}}
  \caption{Wall clock time of parallel~\ac{GBM} with~$N=10^6$ regions
    and overlapping degree~$\alpha=100$. For each value of the number
    of OpenMP threads~$P$, a red dot indicates the number of grid
    cells minimizing the~\ac{WCT}. Each~\ac{WCT} is the average of
    $50$ measurements.}\label{fig:wct-grid}
\end{figure}

We have said above that the optimum number of grid cells in the
parallel~\ac{GBM} algorithm depends on the number of OpenMP
threads~$P$. Figure~\ref{fig:wct-grid} shows this dependency for a
scenario with~$N=10^6$ regions and overlapping degree~$\alpha=100$.  In
the figure we report the~\ac{WCT} as a function of~$P$ and of the
number of grid cells; for each value of~$P$ we put a red dot on the
combination of parameters that provides the minimum~\ac{WCT}. As we
can see, the optimum number of grid cells changes somewhat erratically
as~$P$ increases, although it shows a clear trend suggesting that a
larger number of cells is better for low values of~$P$, while a small
number of cells is better for high values of~$P$. A more precise
characterization of the~\ac{WCT} of the parallel~\ac{GBM} algorithm
would be an interesting research topic, that however falls outside the
scope of the present paper.

% pITM vs. pSBM (scalability)
\begin{figure*}[t]
  \centering
  \subfigure[\label{fig:wct-n}]{\includegraphics[width=.45\textwidth]{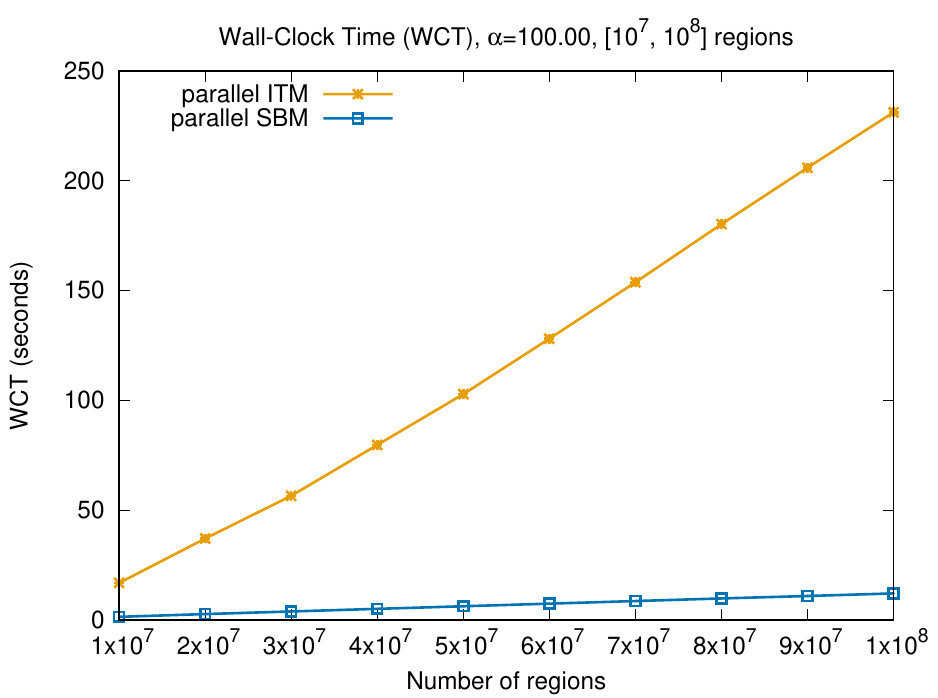}}\quad
  \subfigure[\label{fig:wct-alpha}]{\includegraphics[width=.45\textwidth]{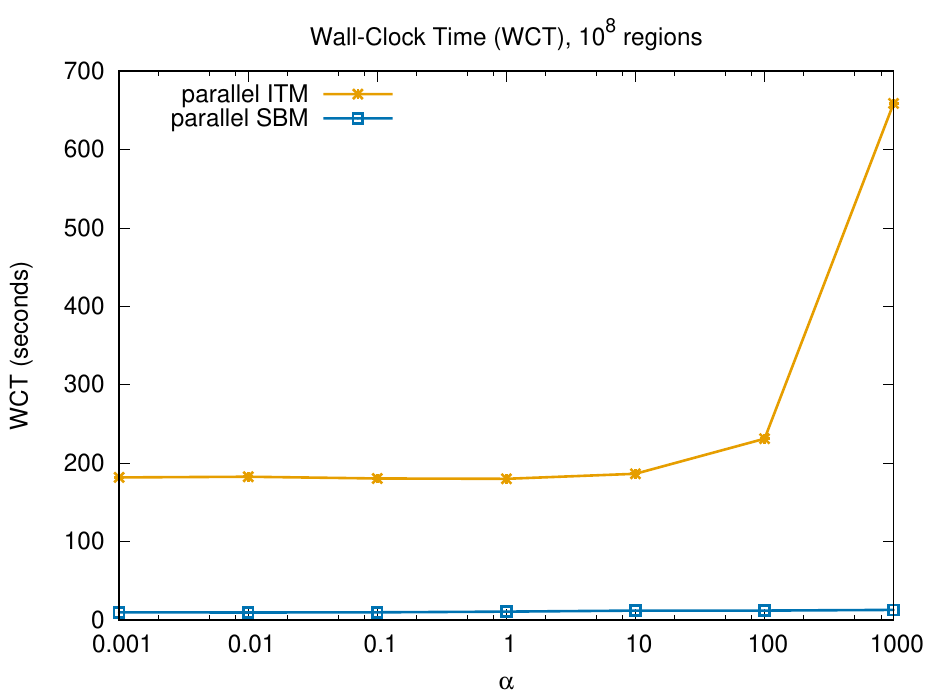}}
  \caption{Wall clock time of parallel~\ac{ITM} and \ac{SBM} as a
    function of the number of regions $N$~(a) and overlapping degree
    $\alpha$~(b) (logarithmic horizontal
    scale).}\label{fig:wct-n-alpha}
\end{figure*}

We now turn our attention on how the~\ac{WCT} changes as a function of
the number of regions~$N$, and as a function of the overlapping
degree~$\alpha$. We set the number of OpenMP threads to $P=32$, the
number of logical cores provided by the test machine.
Figure~\ref{fig:wct-n} shows the~\ac{WCT} of parallel~\ac{ITM}
and~\ac{SBM} for $\alpha = 100$, by varying the number of regions~$N$
in the range $[10^7, 10^8]$. In this range both~\ac{BFM} and~\ac{GBM}
require a huge amount of time, that is orders of magnitude higher than
those of~\ac{ITM} and~\ac{SBM}, and will therefore be omitted from the
comparison. From the figure we observe that the execution times of
both~\ac{ITM} and~\ac{SBM} grow polylogarithmically with~$N$,
supporting the asymptotic analysis in
Section~\ref{sec:parallel-sort-matching}; however, parallel \ac{SBM}
is faster than \ac{ITM}, suggesting that its asymptotic cost has
smaller constant factors.

In Figure~\ref{fig:wct-alpha} we report the \ac{WCT} as a function
of~$\alpha$, for a fixed~$N=10^8$. We observe that, unlike~\ac{ITM},
the execution time of~\ac{SBM} is essentially independent from the
overlapping degree.

% memory usage (peak resident set size, VmHWM)
\begin{figure*}[t]
  \centering%
  \subfigure[\label{fig:memory-extents}]{\includegraphics[width=.45\textwidth]{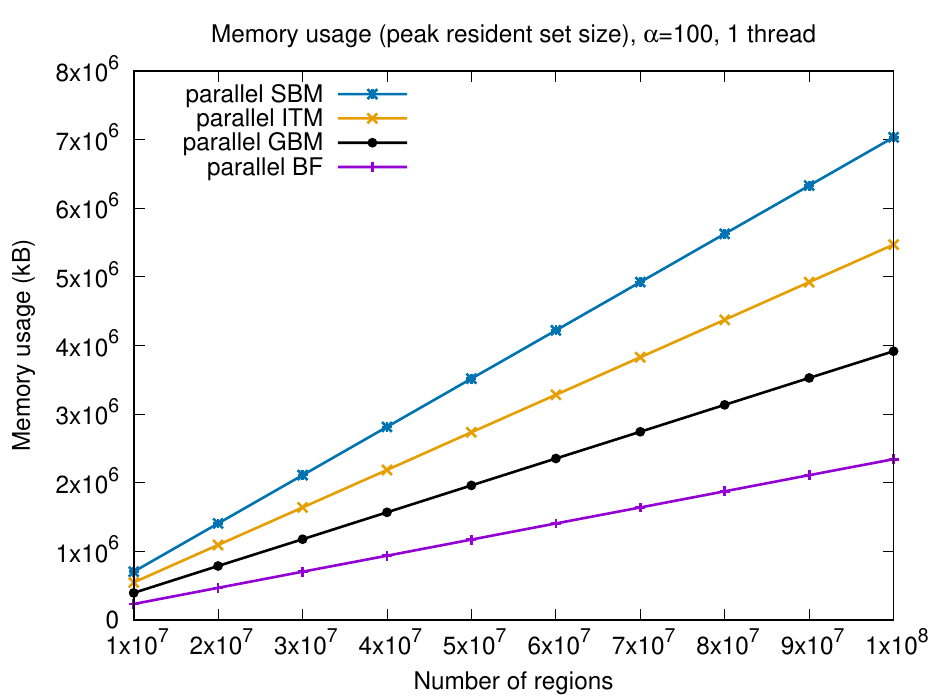}}\quad
  \subfigure[\label{fig:memory-threads}]{\includegraphics[width=.45\textwidth]{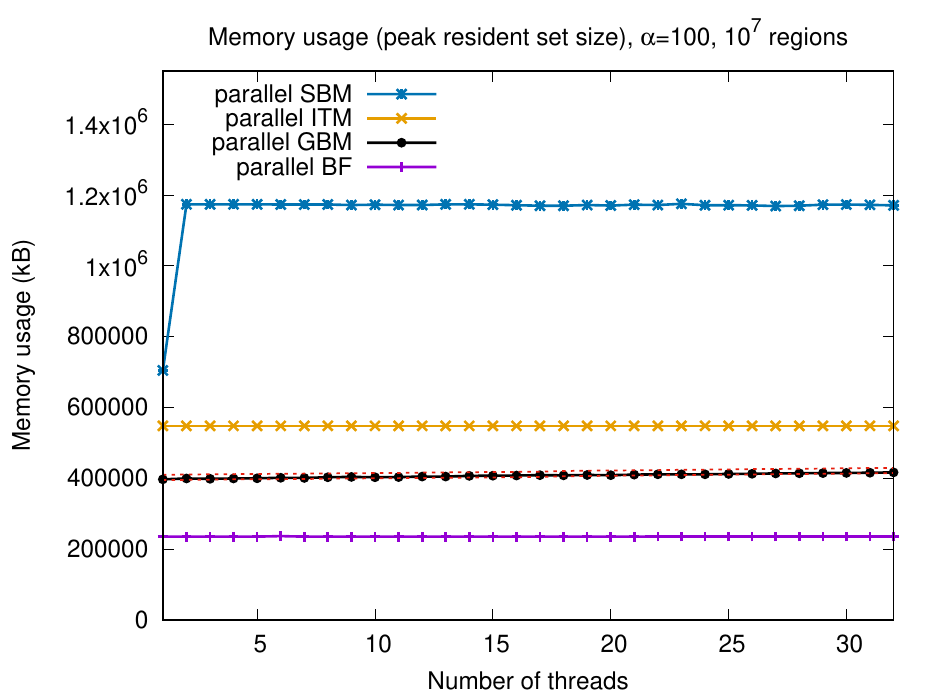}}
  \caption{Memory usage (peak resident set size, VmHWM) of
    parallel~\ac{BFM}, \ac{GBM}, \ac{ITM} and \ac{SBM} with an
    increasing number of regions~(a) or threads~(b), overlapping
    degree $\alpha=100$.}\label{fig:memory}
\end{figure*}

\paragraph*{Memory Usage}
We conclude our experimental evaluation with an assessment of the
memory usage of the parallel~\ac{BFM}, \ac{GBM}, \ac{ITM} and~\ac{SBM}
algorithms. Figure~\ref{fig:memory} shows the peak~\ac{RSS} of the
four algorithms as a function of the number of regions~$N$ and OpenMP
threads~$P$, respectively. The~\ac{RSS} is the portion of a process
memory that is kept in~RAM. Care has been taken to ensure that all
experiments reported in this section fit comfortably in the main
memory of the available machines, so that the~\ac{RSS} represents an
actual upper bound of the amount of memory required by the
algorithms. Note that the data reported in Figure~\ref{fig:memory}
includes the code for the test driver and the input arrays of
intervals.

Figure~\ref{fig:memory-extents} shows that the resident set size grows
linearly with the number of regions~$N$ for all algorithms. \ac{BFM}
has the smaller memory footprint, which is expected since it requires
a small bounded amount of additional memory for a few local variables.
On the other hand, \ac{SBM} requires the highest amount of memory of the
four algorithms, since it allocates larger data structures, namely the
list of endpoints to be sorted, and a few arrays of sets that are used
during the scan phase. In our tests, \ac{SBM} requires approximately
$7$~GB of memory to process~$N=10^8$ intervals, about three times the
amount of memory required by~\ac{BFM}.

In Figure~\ref{fig:memory-threads} we see that the~\ac{RSS} does not
change as the number of OpenMP threads~$P$ increases, with overlapping
degree~$\alpha = 100$ and~$N=10^6$ regions. The anomaly shown
by~\ac{SBM} going from~$P=1$ to~$P=2$ is due to the OpenMP runtime
system.

% pGBM vs. pITM vs. pSBM with the Koln Dataset
\begin{figure*}[t]
  \centering%
  \subfigure[Wall-clock time\label{fig:wct-koln}]{\includegraphics[width=.45\textwidth]{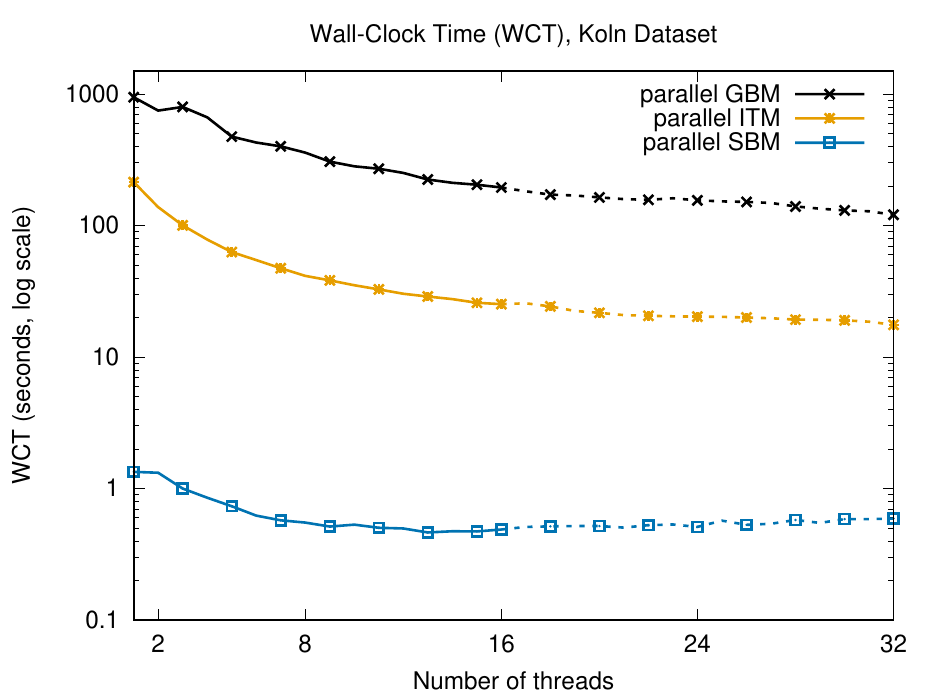}}\quad
  \subfigure[Speedup\label{fig:speedup-koln}]{\includegraphics[width=.45\textwidth]{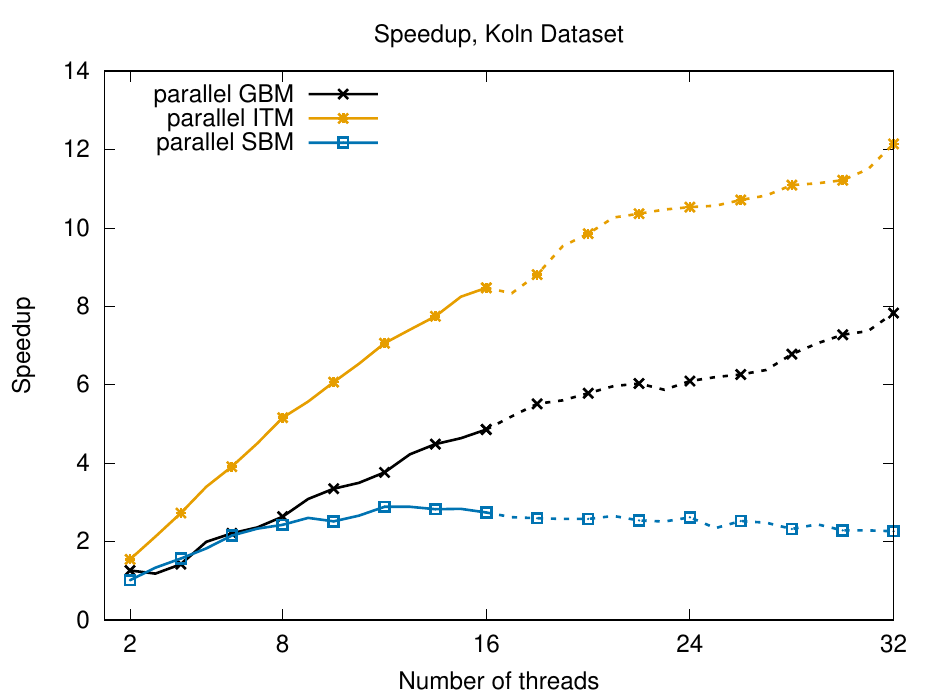}}
  \caption{Wall clock time and speedup of
    parallel~\{~\ac{GBM}, \ac{ITM},~\ac{SBM}\} with the Koln dataset; the~\ac{GBM}
    algorithm uses~$3000$ grid cells. Dashed lines indicate the region
    where the number of OpenMP threads exceeds the number of
    (physical) CPU cores.}\label{fig:wct-speedup-koln}
\end{figure*}

\paragraph*{Performance Evaluation with the Koln Dataset}
So far we have evaluated the~\ac{DDM} implementations using a
synthetic workload. We now complement the analysis by considering a
more realistic workload taken from the vehicular mobility research
domain. Specifically, we use the Cologne dataset~\cite{koln}, a
realistic (although synthetic) trace of car traffic in the city of
Cologne, Germany. The complete
dataset\footnote{\url{http://kolntrace.project.citi-lab.fr/koln.tr.bz2},
  accessed on 2019-03-29} contains the timestamped positions of more
than $700.000$ vehicles moving on the greater urban area of Cologne
($400$ square kilometers) over a period of~$24$ hours.

We consider a portion of the dataset that includes~$541,222$
positions. The~$x$ coordinate of each position is used as the center
of one subscription and one update region; therefore, there are about
$N=10^6$ regions overall. The width of each region is set to~$100$
meters, resulting in approximately~$3.9 \times 10^9$ intersections.

Figure~\ref{fig:wct-speedup-koln} shows the~\ac{WCT} and speedup of
the parallel versions of the~\ac{GBM}, \ac{ITM} and~\ac{SBM}
algorithms; for~\ac{GBM} we used~$3000$ grid cells. As usual, each
data point is the average of~$50$ independent runs. We observe that
the parallel~\ac{GBM} algorithm is the slowest of the three, while
parallel~\ac{SBM} is the fastest by a wide margin (three orders of
magnitude faster than~\ac{SBM}, two orders of magnitude faster
than~\ac{ITM}). Since~\ac{SBM} is very fast on this benchmark, its
poor scalability is caused by the parallelization overhead of OpenMP
that has a higher impact on low wall-clock times.

%%%%%%%%%%%%%%%%%%%%%%%%%%%%%%%%%%%%%%%%%%%%%%%%%%%%%%%%%%%%%%%%%%%%%%%%%%%%%%
\section{Conclusions}\label{sec:conclusions}

In this paper we described and analyzed two parallel algorithms for
the region matching problem on shared-memory architectures. The region
matching problem consists of enumerating all intersections among two
sets of subscription and update regions; a region is a
$d$-dimensional, iso-oriented rectangle. The region matching problem
is at the core of the~\acl{DDM} service which is part of
the~\acl{HLA}.

The region matching problem in~$d$ dimensions can be reduced to the
simpler problem of computing intersections among one-dimensional
segments. The first parallel solution to the 1D matching problem,
called~\ac{ITM}, is based on an interval tree data structure. An
interval tree is a binary balanced search tree that can store a set of
segments, and can be used to efficiently enumerate all intersections
with a given query segment. Once built, an interval tree can be
efficiently queried in parallel. The second solution is based on a
parallel extension of~\ac{SBM}, a state-of-the-art solution to
the~\ac{DDM} problem.

We have implemented the parallel versions of~\ac{ITM} and~\ac{SBM}
using the C/C++ programming languages with OpenMP extensions. These
algorithms have been compared with parallel implementations of the
Brute-Force and Grid-Based matching algorithms. The results show that
the parallel versions of~\ac{ITM} and~\ac{SBM} are orders of magnitude
faster than (the parallel versions of) Brute-Force and Grid-Based
matching. Among the four algorithms considered, parallel~\ac{SBM} is
the fastest in all scenarios we have examined. The~\ac{ITM} algorithm,
while slower than~\ac{SBM}, can be easily extendable to cope with
dynamic regions since the interval tree allows efficient insertion and
deletion of regions. In fact, a version of~\ac{SBM} that can
efficiently handle region updates has already been
proposed~\cite{Pan2011}, but it can not be readily adapted to the
parallel version of~\ac{SBM} discussed in this paper. Developing a
parallel and dynamic version of~\ac{SBM} is the subject of ongoing
research.

In this paper we focused on shared-memory architectures, i.e.,
multicore processors, since they are virtually ubiquitous and well
supported by open, standard programming frameworks (OpenMP).
Modern~\acp{GPU} can be faster than contemporary~CPUs, and are
increasingly being exploited for compute-intensive applications.
Unfortunately, the parallel~\ac{ITM} and~\ac{SBM} algorithms presented
in this paper are ill-suited for implementations on \acp{GPU}, since
they rely on data structures with irregular memory access patterns
and/or frequent branching within the code. Reworking the
implementation details of~\ac{ITM} and~\ac{SBM} to better fit the
\emph{Symmetric Multithreading} model of modern~\acp{GPU} is still and
open problem that requires further investigation.

\appendix

\section{Notation}

\begin{tabularx}{\textwidth}{lX}
  $\mathbf{S}$ & Subscription set $\mathbf{S} = \{S_1, \ldots, S_n\}$\\
  $\mathbf{U}$ & Update set $\mathbf{U} = \{U_1, \ldots, U_m\}$\\
  $n$          & Number of subscription regions\\
  $m$          & Number of update regions\\
  $N$          & Number of subscription and update regions ($N = n + m$)\\
  $K_s$        & N. of intersections of subscription region $s$ with \newline all update regions  ($0 \leq K_s \leq m$)\\
  $K_u$        & N. of intersections of update region $u$ with \newline all subscription regions ($0 \leq K_u \leq n$)\\
  $K$          & Total number of intersections ($0 \leq K \leq n \times m$)\\
  $\alpha$     & Overlapping degree ($\alpha > 0$)\\
  $P$          & Number of processors\\
\end{tabularx}

\section{Acronyms}

\begin{acronym}[NUMA]
\acro{BFM}{Brute Force Matching}
\acro{DDM}{Data Distribution Management}
\acro{GBM}{Grid Based Matching}
\acro{GPU}{Graphics Processing Unit}
\acro{HLA}{High Level Architecture}
\acro{HT}{Hyper-Threading}
\acro{ITM}{Interval Tree Matching}
\acro{NUMA}{Non Uniform Memory Access}
\acro{OMT}{Object Model Template}
\acro{OS}{Operating System}
\acro{RSS}{Resident Set Size}
\acro{RTI}{Run-Time Infrastructure}
\acro{SBM}{Sort-based Matching}
\acro{SIMD}{Single Instruction Multiple Data}
\acro{STL}{Standard Template Library}
\acro{UMA}{Uniform Memory Access}
\acro{WCT}{Wall Clock Time}
\end{acronym}

\bibliographystyle{ACM-Reference-Format}
\bibliography{parallel-ddm}

\end{document}